\documentclass[12pt]{article}
 
\usepackage{bm}
\usepackage[
backend=biber,
style=apa,
natbib=true,
]{biblatex}
\usepackage{booktabs}
\usepackage{mathtools} 
\usepackage{amssymb} 
\usepackage{xcolor}
\usepackage[margin=1in]{geometry}
\usepackage{microtype} 
\usepackage{lineno} 
\usepackage{subcaption} 
\usepackage{graphicx}
\usepackage{varwidth} 
\usepackage{hyperref} 

\hypersetup{%
  colorlinks = true, 
  citecolor = blue,
  urlcolor = blue,
  bookmarks = true 
}

\newcommand{\inst}[1]{\vspace{1pt} \unskip\(^{#1}\)}

\begin{document}

\title{\vspace{-2cm}Modelling sub-daily precipitation extremes with the blended generalised extreme value
  distribution
}

\author{Silius M. Vandeskog\inst{1} \and
  Sara Martino\inst{1} \and
  Daniela Castro-Camilo\inst{2} \and
  Håvard Rue\inst{3}
}

\date{}

\maketitle

\noindent \inst{1} Norwegian University of Science and Technology (NTNU) \\
\inst{2} University of Glasgow \\
\inst{3} King Abdullah University of Science and Technology (KAUST)

\begin{abstract}
  A new method is proposed for modelling the yearly maxima of sub-daily precipitation, with the aim
  of producing spatial maps of return level estimates. Yearly precipitation maxima
  are modelled using a Bayesian hierarchical model with a latent Gaussian field, with the 
  blended generalised
  extreme value (bGEV) distribution used as a substitute for the more standard generalised
  extreme value (GEV) distribution. Inference is made less wasteful
  with a novel two-step procedure that performs separate modelling of the scale parameter of the
  bGEV distribution using peaks over threshold data. Fast inference is performed using
  integrated nested Laplace approximations (INLA) together with the stochastic partial differential
  equation (SPDE) approach, both implemented in \texttt{R-INLA}. Heuristics for improving the
  numerical stability of \texttt{R-INLA} with the GEV and bGEV distributions are also presented.
  The model is fitted to yearly maxima of sub-daily precipitation from the south of Norway, and is
  able to quickly produce high-resolution return level maps with uncertainty. The proposed
  two-step procedure provides an improved model fit over standard inference techniques when
  modelling the yearly maxima of sub-daily precipitation with the bGEV distribution.
\end{abstract}


\section{Introduction}
\label{sec:introduction}

Heavy rainfall over short periods of time can cause flash floods, large economic losses and immense
damage to infrastructure. The World Economic Forum states that climate action failure
and extreme weather events are perceived among the most likely and most impactful global risks in
2021 \citep{wef_2021}. Therefore, a better understanding of heavy rainfall can be of utmost
importance for many decision-makers, e.g.,\ those that are planning the construction or maintenance
of important infrastructure. In this paper, we create spatial maps with estimates of
large return levels for sub-daily precipitation in Norway.
Estimation of return levels is best described within the framework
of extreme value theory, where the most common methods are the block maxima  and the peaks
over threshold
 \citep[e.g.][]{davison15_statis_extrem,coles01_introd_statis_model_extrem_values}. Due to
low data quality (see Section~\ref{sec:data} for more details) and the difficulty of selecting
high-dimensional thresholds, we choose to use
the block maxima method for estimating the precipitation return levels. This method
is based on modelling the maximum of a large block of random variables with the generalised extreme
value (GEV) distribution, which is the only non-degenerate limit distribution for a standardised
block maximum \citep{fisher28_limit_forms_frequen_distr_larges}. When working with environmental
data, blocks are typically chosen to have a size of one year
\citep{coles01_introd_statis_model_extrem_values}. Inference with the GEV
distribution is difficult, partially because its support  depends on its
parameter values. \citet{castro-camilo21_pract_strat_gev_regres_model_extrem} propose to ease
inference by substituting the
GEV distribution with the blended generalised extreme value (bGEV) distribution, which has the right
tail of a Fréchet distribution and the left tail of a Gumbel distribution, resulting in a heavy-tailed
distribution with a
parameter-free support. Both \citet{castro-camilo21_pract_strat_gev_regres_model_extrem} and
\citet{vandeskog21_model_block_maxim} demonstrate with simulation studies that the bGEV 
distribution performs well as a substitute for the GEV distribution when estimating properties
of the right tail. Additionally, in this paper we develop a simulation study that shows how
the parameter-dependent support of the GEV distribution can lead to numerical problems during
inference, while inference with the bGEV distribution is more robust. This can be of crucial
importance in complex and high-dimensional settings, and consequently we choose to model the
yearly maxima of sub-daily precipitation using the bGEV distribution.

Modelling of extreme daily precipitation has been given much attention in the literature, and it is
well established that precipitation is a heavy-tailed phenomenon
\citep[e.g.][]{wilson05_fundam_probab_distr_heavy_rainf, katz02_statis_extrem_hydrol,
  papalexiou13_battl_extrem_value_distr},
which makes the bGEV distribution a possible model for yearly precipitation
maxima.
Spatial modelling of extreme daily precipitation has also received a great amount of
interest.
\citet{cooley07_bayes_spatial_model_extrem_precip_retur_level} combine Bayesian
hierarchical modelling with a generalised Pareto likelihood for
estimating large return values for daily precipitation. Similar methods are also applied
by \citet{sang09_hierar_model_extrem_values_obser, geirsson15_comput_effic_spatial_model_annual,
  davison12_statis_model_spatial_extrem, opitz18_inla_goes_extrem}, using either the block maxima
or the peaks over threshold approach.
Using a multivariate peaks over threshold approach,
\citet{castro-camilo20_local_likel_estim_compl_tail} propose local likelihood inference for a 
specific factor copula model to deal with complex
non-stationary dependence structures of precipitation over the contiguous
U.S. Spatial modelling of extreme sub-daily precipitation is
more difficult, due to less available data sources. Consequently, this is often performed using
intensity-duration-frequency relationships where one pools together information from multiple
aggregation times in order to estimate return levels
\citep{koutsoyiannis98_mathem_framew_study_rainf_inten, ulrich20_estim_idf_curves_consis_over,
  lehmann16_spatial_model_framew_charac_rainf, wang16_bayes_hierar_model_spatial_extrem}.
Spatial modelling of extreme hourly precipitation in Norway has previously been performed by
\citet{dyrrdal15_bayes_hierar_model_extrem_hourl_precip_norway}. After their work was published,
the number of observational sites for hourly precipitation in Norway has greatly increased.
We aim to improve their return level estimates by including all the new data
that have emerged over the last years.
We model sub-daily precipitation using a spatial Bayesian hierarchical model with a bGEV
likelihood and a latent Gaussian field. In order to keep our model simple, we do not pool together information
  from multiple aggregation times, making our model purely spatial.
The model assumes conditional independence between observations, which makes it able to
estimate the marginal distribution of
extreme sub-daily precipitation at any location, but unable to successfully estimate joint
distributions over multiple locations. In the case of hydrological processes such as precipitation,
ignoring dependence might lead to an underestimation of the risk of flooding. However,
\citet{davison12_statis_model_spatial_extrem} find that models where the response variables are
independent given some latent process can be a good choice when the
aim is to estimate a spatial map of marginal return levels.

High-resolution spatial modelling can demand a lot of
computational resources and be highly time-consuming. The framework of integrated nested Laplace
approximations \citep[INLA;][]{rue09_approx_bayes_infer_laten_gauss} allows for a considerable
speed-up by using numerical approximations instead of sampling-based inference methods like Markov
chain Monte Carlo (MCMC). Inference with a spatial Gaussian latent field can be even further sped up
with the so-called stochastic partial differential equation
\citep[SPDE;][]{lindgren11_explic_link_between_gauss_field} approach of representing a Gaussian 
random field using a Gaussian Markov random field that is the approximate solution of a specific
SPDE. Both INLA and the SPDE approach have been
implemented in the \texttt{R-INLA} library, which is used for performing inference with our
model \citep{bivand2015spatial, rue17_bayes_comput_with_inla,
  bakka18_spatial_model_with_r_inla}. \texttt{R-INLA} requires a log-concave likelihood to
ensure numerical stability during inference. However, neither the GEV likelihood nor the bGEV
likelihood are log-concave, which can cause inferential issues. We present heuristics
for mitigating the risk of numerical instability caused by a lack of log-concavity.

A downside of the block maxima method is that inference can be somewhat wasteful compared
to the peaks over threshold method. Additionally, most of
the available weather stations in Norway that measure hourly precipitation are young and contain
quite short time series.
This data sparsity makes it challenging to place complex models on the parameters of the bGEV
distribution in the hierarchical model. A promising method of accounting for data-sparsity is the recently
developed sliding block estimator, which allows for better data utilisation by not requiring that
the block maxima used for inference come from disjoint blocks
\citep{buecher18_infer_heavy_tailed_station_time,zou19_multip_block_sizes_overl_block}. However, to
the best of our knowledge, no theory has yet been developed for using the disjoint block estimator
on non-stationary time series, or for performing Bayesian inference with the disjoint block
estimator. \citet{vandeskog21_model_block_maxim} propose a
new two-step procedure that allows for less wasteful and more stable inference with the block maxima
method by separately modelling the scale parameter of the bGEV distribution using peaks over
threshold data. Having modelled the scale parameter, one can standardise the block maxima so the
scale parameter can be considered as a constant, and then estimate the remaining bGEV parameters. 
\citet{buecher21_horse_race_between_block_maxim} suggests that, when modelling stationary time
series, the peaks over threshold technique is preferable over block maxima if the interest
lies in estimating large quantiles of the stationary distribution of the times series. The opposite holds if the
interest lies in estimating return levels, i.e.\ quantiles of the distribution of the block
maxima. Thus, both methods have different strengths, and by
using this two-step procedure, one can take advantage of the merits and improve
the pitfalls of both methods. 
We apply the two-step procedure for modelling sub-daily precipitation and compare the performance
with that of a standard block maxima model where all the bGEV parameters are estimated jointly.

The remainder of the paper is organised as follows. Section~\ref{sec:data} introduces the hourly
precipitation data and all explanatory variables used for modelling. Section~\ref{sec:method}
presents the bGEV distribution and describes the Bayesian hierarchical model along with the two-step
modelling procedure. Additionally, heuristics for improving the numerical stability of \texttt{R-INLA} are
proposed, and a score function for evaluating model performance is presented. In
Section~\ref{sec:case-study} we perform modelling of the yearly precipitation maxima in Norway. A
cross-validation study is performed for evaluating the model fit, and a map of return levels is
estimated. Conclusions are presented in Section~\ref{sec:conclusion}.

\section{Data}
\label{sec:data}

\subsection{Hourly precipitation data}
\begin{figure}
  \centering
  \includegraphics[width=.8\linewidth]{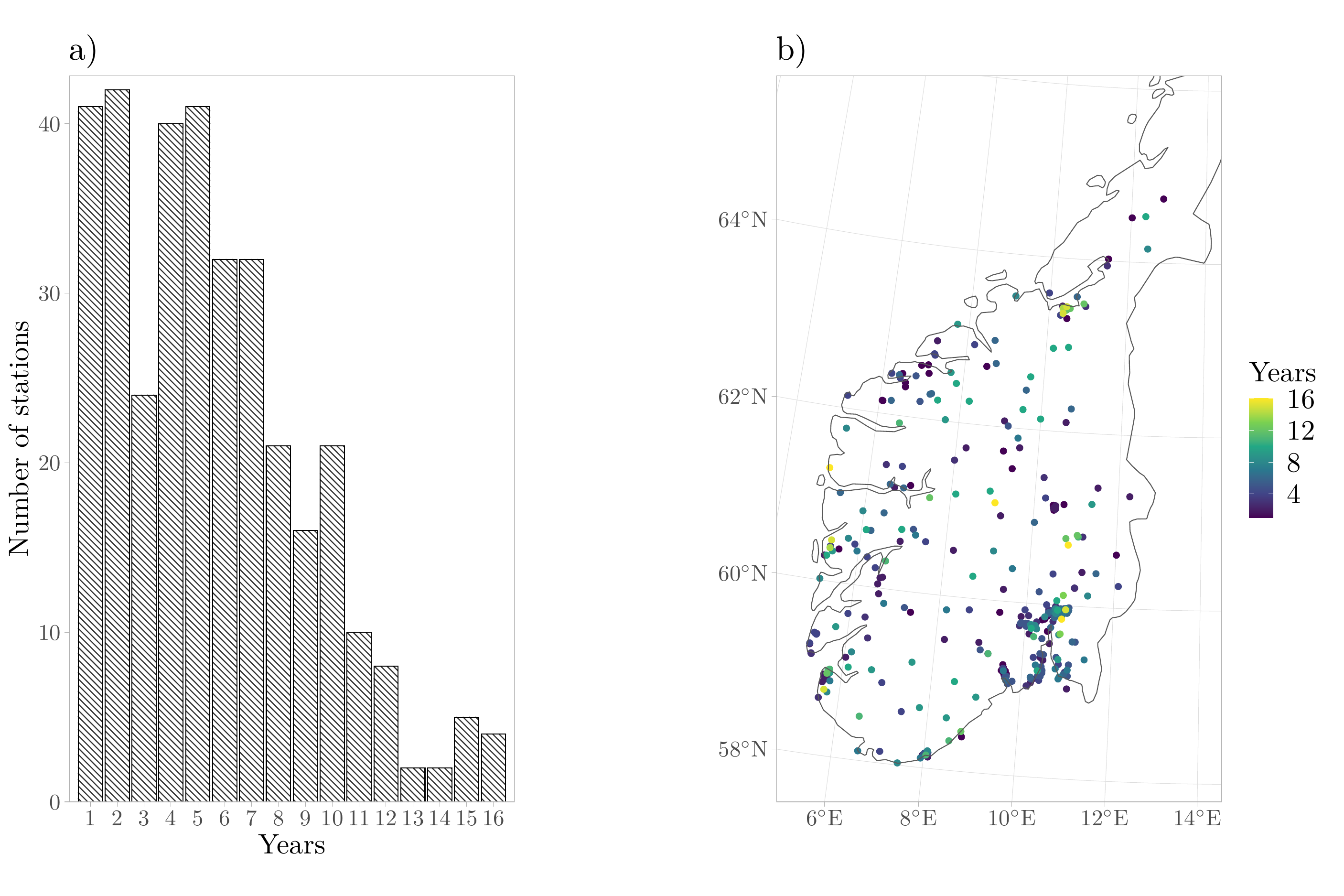}
  \caption{a) A histogram displaying the number of usable yearly precipitation maxima for all the
    weather stations used in this paper. b) The location of the 341 weather stations. The number of usable
    yearly precipitation maxima from each station are displayed using different
    colours. Note that some points overlap in areas with high station densities.}
  \label{fig:station-locations}
\end{figure}

Observations of hourly aggregated precipitation from a total of 380 weather stations
in the south of Norway are downloaded from an open archive of historical weather data from MET
  Norway (\url{https://frost.met.no}). The oldest weather 
stations contain observations from 1967, but approximately 90 percent of the available weather
stations are established after 2000. Each observation comes with a quality code, but almost all
observations from before 2005 are of unknown quality. An inspection of the time series with unknown
quality detects unrealistic precipitation observation ranging from \(-300\)~mm/hour to
\(400\)~mm/hour. Other unrealistic patterns, like \(50\)~mm/hour precipitation for more than three
hours in a row, or no precipitation during more than half a year, are also detected. The data set
contains large amounts of missing data, but these are often recorded as 0~mm/hour, instead of being
recorded as missing. Thus, there is no way of
knowing which of the zeros represent missing data and which represent an hour without
precipitation. Having detected all of this, we decide to remove all observations with unknown or bad
quality flags, which accounts for approximately \(14\%\) of the total number of observations.
Additionally, we clean the data by removing all observations
from years with more than \(30\%\) missing data and from years where more
than two months contain less than \(20\%\) of the possible observations. This data cleaning is
performed to increase the 
probability that our observed yearly maxima are close or equal to the true yearly maxima.
Having cleaned the data, we
are left with \(72\%\) of the original observations, distributed over 341 weather stations and
spanning the years 2000 to 2020. The total number of usable yearly maxima is
approximately 1900. Figure~\ref{fig:station-locations} displays the distribution of the number of
usable yearly precipitation maxima per weather station. The majority of the weather stations contain
five or less usable yearly maxima, and approximately 50 stations have more than 10 usable maxima.
Figure~\ref{fig:station-locations} also displays the location of all the weather stations. A
large amount of the stations are located close to each other, in the southeast of Norway. Such
spatial clustering can be an indicator for preferential sampling. However, we do not believe that
preferential sampling is an issue for our data. The weather stations are mostly placed in
locations with high population densities, and to the best of our knowledge there is no strong
dependency between population density and extreme precipitation in Norway, as there are 
large cities located both in dry and wet areas of the country.
Even though most stations are located in areas with high population densities, there is still a good 
spatial coverage of the entire area of interest, also for areas with low population densities.

The yearly maxima of precipitation accumulated over \(1, 2, \ldots, 24\) hours
are computed for all locations and available years. A rolling window approach with a step size of 1
hour is used for locating the precipitation maxima. As noted by
\citet{robinson00_extrem_analy_proces_sampl_at_differ_frequen}, a sampling frequency of one hour
is not enough to observe the exact yearly maximum of hourly precipitation. With this sampling
frequency, one only observes the precipitation during the periods 00:00-01:00, 01:00-02:00,
etc.,\ whereas the maximum precipitation might occur e.g.\ during the period
14:23-15:23. Approximately
half of the available weather stations have a sampling frequency of one minute, while the other half
only contain hourly observations. We therefore use a sampling
frequency of one hour for all weather stations, as this allows us to use all the 341 weather
stations without having to account for varying degrees of sampling frequency bias in our model.

\citet{dyrrdal15_bayes_hierar_model_extrem_hourl_precip_norway} used the same data source for
estimating return levels of hourly precipitation. They fitted their models to hourly precipitation
maxima using only 69 weather stations from all over Norway.
However, they received a cleaned data set from the Norwegian Meteorological Institute,
resulting in time series with lengths up to 45 years. Our data cleaning approach is more
strict than that of \citet{dyrrdal15_bayes_hierar_model_extrem_hourl_precip_norway} in the sense
that it results in shorter time series by removing all data of uncertain
quality. On the other hand, we include more locations and get a considerably better spatial
coverage, by keeping all time series with at least one good year of observations. 

The main focus of this paper is the novel methodology for fast and accurate estimation of
return levels, and we believe 
that we have prepared the data well enough to give a good demonstration of our proposed model and to
achieve reliable return level estimates for sub-daily precipitation. It is trivial to add more, or
differently cleaned data, to improve the return level estimates at a later time.

\subsection{Explanatory variables}

We use one climate-based and four orographic explanatory variables. These are displayed
in Table~\ref{tab:covariates}. Altitude is extracted from a digital elevation model of resolution \(50
\times 50\) m\(^2\), from the Norwegian Mapping Authority (\url{https://hoydedata.no}). The distance
to the open sea is computed using the digital elevation model. Precipitation climatologies for the
period 1981-2010 are modelled
by \citet{crespi18_high_resol_month_precip_climat_over_norway}. The climatologies do
not cover the years 2011-2020, from which most of the observations come. We assume that the
precipitation patterns have not changed overly much and that they are still representative for the
years 2011-2020. \citet{hanssen-bauer98_annual_seasonal_prec_var} find that, in most southern
regions of Norway, the only season with a significant increase in precipitation is Autumn. This
strengthens our assumption that the change in precipitation patterns is slow enough to not be
problematic for us.

\citet{dyrrdal15_bayes_hierar_model_extrem_hourl_precip_norway} include additional explanatory
variables in their model, such as temperature, summer precipitation and the yearly number of
wet days. They find mean summer precipitation to be one of the most important explanatory
variables. We compute these explanatory variables at all station locations using the gridded
seNorge2 data product \citep{lussana18_daily_precip_obser_gridd_datas,
  lussana18_three_dimen_spatial_inter_m}. Our examination finds that yearly
precipitation, summer precipitation and the yearly number of wet days are close to \(90\%\)
correlated with each other. There is also a negative correlation between temperature and altitude of
around -\(85\%\). Consequently, we choose to not use any more explanatory variables for modelling,
as highly correlated variables might lead to identifiability issues during parameter estimation. 

\begin{table}
  \centering
  \renewcommand*{\arraystretch}{1.4}
  \caption{Explanatory variables used for modelling sub-daily precipitation extremes. The
    two rightmost columns show which explanatory variables are used for modelling which parameters
    of the bGEV distribution for yearly precipitation (see Section~\ref{sec:model-framework}).}
  \label{tab:covariates}
  \begin{tabular}{p{5cm}p{7cm}lll}
    \toprule
    Explanatory variable & Description & Unit & \(\textbf{x}_\mu\) & \(\textbf{x}_\sigma\) \\
    \midrule
    Mean annual precipitation &  Mean annual precipitation
                                  for the years 1981-2010 & mm & \checkmark &  \\
    Easting & Eastern coordinate (UTM 32)
                                       & km & \checkmark & \checkmark \\
    Northing & Northern coordinate (UTM 32)
                                       & km & \checkmark & \checkmark \\
    Altitude & Height above sea level & m & \checkmark &  \\
    Distance to the open sea &  Shortest distance to the open sea & km & \checkmark & \checkmark \\
    \bottomrule
  \end{tabular}
\end{table}

\section{Methods}
\label{sec:method}

\subsection{The bGEV distribution}
\label{sec:bgev}

Extreme value theory concerns the statistical behaviour of extreme events, possibly larger
than anything ever observed. It provides a framework where probabilities associated
with these events can be estimated by extrapolating into the tail of the distribution. This can be
used for e.g.\ estimating large quantiles, which is the aim of this
work \citep[e.g.][]{davison15_statis_extrem, coles01_introd_statis_model_extrem_values}.
A common approach in extreme value theory is the block maxima method. Assume that the limiting distribution
of the standardised block maximum \((Y_k - b_k) / a_k\) is non-degenerate, where \(Y_k =
\max\{X_1, X_2, \ldots, X_k\}\) is the maximum over \(k\) random variables from a stationary
stochastic process, and \(\{b_k\}\) and \(\{a_k > 0\}\) are some appropriate sequences of standardising
constants. Then, for large enough block sizes \(k\), the distribution of the block maximum \(Y_k\)
is approximately equal to the GEV distribution with
\citep{fisher28_limit_forms_frequen_distr_larges,coles01_introd_statis_model_extrem_values}
\begin{linenomath*}
  \begin{equation}
    \label{eq:gev}
    P(Y_k \leq y) \approx \begin{cases}
      \exp\left\{-[1 + \xi (y - \mu_k) / \sigma_k]^{-1/\xi}_+\right\}, & \xi \neq 0, \\
      \exp\left\{-\exp[-(y - \mu_k) / \sigma_k]\right\}, & \xi = 0,
    \end{cases},
  \end{equation}
\end{linenomath*}
where \((a)_+ = \max\{a, 0\}\), \(\sigma_k > 0\) and \(\mu_k, \xi \in \mathbb R\). In most
settings, \(k\) is fixed, so we denote \(\sigma = \sigma_k\) and \(\mu = \mu_k\).
A challenge with the GEV distribution is that its support depends on its
parameters. This complicates inference procedures such as maximum likelihood estimation
\citep[e.g.][]{buecher17_maxim_likel_estim_gener_extrem_value_distr,
  smith85_maxim_likel_estim_class_nonreg}, and can be particularly problematic in a 
covariate-dependent setting with spatially varying parameters, as it might also introduce artificial
boundary restrictions such as an unnaturally large lower bound for yearly maximum precipitation.
\citet{castro-camilo21_pract_strat_gev_regres_model_extrem} propose the bGEV distribution as an
alternative to the GEV
distribution in settings where the tail parameter \(\xi\) is non-negative. The support of the bGEV
distribution is parameter-free and infinite. This allows for more numerically stable
inference, while also avoiding the possibility of estimated lower bounds that are larger than future observations.
The bGEV distribution function is
\begin{linenomath*}
\begin{equation}
  \label{eq:bgev}
  H(y; \mu, \sigma, \xi, a, b) = {F(y; \mu, \sigma, \xi)}^{v(y; a, b)} {G(y; \tilde{\mu},
    \tilde{\sigma})}^{1 - v(y; a, b)},
\end{equation}
\end{linenomath*}
where \(F\) is a GEV distribution with \(\xi \geq 0\) and \(G\) is a Gumbel
distribution. The weight function is equal to
\begin{linenomath*}
\begin{equation*}
  \label{eq:bgev-weight}
  v(y; a, b) = F_\beta\left(\frac{y - a}{b - a}; c_1, c_2\right),
\end{equation*}
\end{linenomath*}
where \(F_\beta(\cdot; c_1, c_2)\) is the distribution function of a beta 
distribution with parameters \(c_1 = c_2 = 5\), which leads to a symmetric and computationally
efficient weight function.
The weight \(v(y; a, b)\) is zero for \(y \leq a\) and one for
\(y \geq b\), meaning that the left tail of the bGEV distribution is equal to the left tail in \(G\), while
the right tail is equal to the right tail in \(F\). The choice of the weight \(v(y; a, b)\)
  should not considerably affect inference if we let the difference between \(a\) and \(b\) be small.
The parameters \(\tilde{\mu}\) and
\(\tilde{\sigma}\) are injective functions of \((\mu, \sigma, \xi)\) such that the bGEV distribution
function is continuous and \(F(y; \mu, \sigma, \xi)
= G(y; \tilde{\mu}, \tilde{\sigma})\) for \(y \in \{a, b\}\). Setting \(a = F^{-1}(p_a)\)
and \(b = F^{-1}(p_b)\) with small probabilities \(p_a = 0.1,\ p_b = 0.2\) makes it
possible to model the right tail of the GEV distribution without any of the problems caused by a
finite left tail. See \citet{castro-camilo21_pract_strat_gev_regres_model_extrem} for guidelines
on how to choose \(c_1\), \(c_2\), \(p_a\) and \(p_b\).

In the supplementary material we present a simulation study where both the GEV distribution
and the bGEV distribution are fitted to univariate samples from a GEV distribution. We demonstrate
how a small change in initial values can cause large numerical problems for inference with the
GEV distribution, and no noticeable difference for inference with the bGEV distribution.
The fact that considerable numerical problems can arise for the GEV distribution in a univariate
setting with large sample sizes and perfectly GEV-distributed data strongly indicates that the GEV
distribution is not robust enough to be used reliably in complex, high-dimensional problems with
noisy data. The bGEV distribution is more robust than the GEV distribution, and we therefore prefer it
over the GEV distribution for modelling precipitation maxima in Norway.

Although the bGEV distribution is more robust than the GEV distribution, it might still
seem unnatural to model block maxima using the bGEV distribution, when it is known that
the correct limiting distribution is the GEV distribution. However, we argue that the bGEV
would be a good choice for modelling heavy tailed block maxima even if it had not been more robust
than the GEV distribution.
In multivariate extreme value theory it is common to
assume that the tail parameter \(\xi\) of the GEV distribution is constant in time and/or space
\citep[e.g.][]{opitz18_inla_goes_extrem, koutsoyiannis98_mathem_framew_study_rainf_inten,
  castro-camilo19_splic_gamma_gener_paret_model, sang10_contin_spatial_proces_model_spatial_extrem_values}.
This assumption is often made, not because one truly believes that it should be constant, but
because estimation of \(\xi\) is difficult, and models with a constant
\(\xi\) often are ``good enough''. The tail parameter is incredibly important for the shape of the
GEV distribution, and small changes in \(\xi\) can lead to large changes in return levels, and
even affects the existence of distributional moments. A model where \(\xi\) varies in space can
therefore e.g.\ provide model fits with a finite mean in one location and an infinite mean
in the neighbouring location. Such a model can also give scenarios where a new observation at one
location can change the existence of moments in other, possibly far away, locations.
Thus, even though it might seem unnatural to use a constant tail parameter, these
models often provide more natural fits to data than the models that allow \(\xi\)
to vary in space. We claim that the bGEV distribution fulfils a similar role as a model with
constant \(\xi\), but for the model
support instead of the moments. When \(\xi\) is positive, the support of
the GEV distribution varies with its parameter values. In regression settings with covariates and
finite amounts of data one can therefore experience unnatural lower bounds that are known to be
wrong. Furthermore, if only one new observation is smaller than the estimated lower limit, the
entire model fit will be invalidated. We therefore prefer the bGEV, which completely removes the lower
bound while still having the right tail of the GEV distribution, thus yielding a model that is
``good enough'' for estimating return levels, but without the unwanted model properties in the
left tail of the GEV distribution.

Naturally, the bGEV distribution can only be applied for modelling exponential- or heavy-tailed phenomena \((\xi
\geq 0)\). However, it is well established that extreme precipitation should be modelled with a
non-negative tail parameter.
\citet{cooley07_bayes_spatial_model_extrem_precip_retur_level} performs
Bayesian spatial modelling of extreme daily precipitation in Colorado and find that the tail
parameter is positive and less than 0.15. \citet{papalexiou13_battl_extrem_value_distr} examine more
than 15000 records of daily precipitation worldwide and conclude that the Fréchet distribution
performs the best. They propose that even when the data suggests a negative tail parameter,
it is more reasonable to use a Gumbel or Fréchet distribution.
Less information is available concerning the distribution of extreme
sub-daily precipitation. However, \citet{koutsoyiannis98_mathem_framew_study_rainf_inten}
argues that the distribution of precipitation should not have an upper bound for any aggregation
period, so \(\xi\) must be 
non-negative. \citet{van12_spatial_regres_model_extrem_precip_belgium} estimate the distribution
of yearly precipitation maxima in Belgium for aggregation times down to 1 minute, and find that
the estimates of \(\xi\) increase as the aggregation times decreases,
meaning that the tail parameter for sub-daily precipitation should be larger than
for daily precipitation. 
\citet{dyrrdal16_estim_extrem_areal_precip_norway} estimate \(\xi\) for
daily precipitation in Norway from the seNorge1 data
product \citep{tveito05_gis_based_agro_ecolog_decis, mohr09_compar_version} and conclude that the
tail parameter estimates are non-constant in space
and often negative. However, the authors do not provide confidence intervals or p-values and do
not state whether the estimates are significantly different from zero.
Based on our own exploratory analysis (results not shown) and the overwhelming evidence in the
literature, we assume that sub-daily precipitation is a heavy-tailed phenomenon.
 
Following \citet{castro-camilo21_pract_strat_gev_regres_model_extrem}, we reparametrise the bGEV
distribution from \((\mu, \sigma, \xi)\) to 
\((\mu_\alpha, \sigma_\beta, \xi)\), where the location parameter \(\mu_\alpha\) is equal to the \(\alpha\)
quantile of the bGEV distribution if \(\alpha \geq p_b\). The scale parameter \(\sigma_\beta\), hereby
denoted the spread parameter, is equal to the difference between the \(1 - \beta/2\) quantile and
the \(\beta/2\) quantile of the bGEV distribution if \(\beta / 2 \geq p_b\). There is a
one to one relationship between the new and the old parameters.
The new parametrisation is advantageous as it is considerably easier to
interpret than the old parametrisation. The parameters \(\mu_\alpha\) and \(\sigma_\beta\) are
directly connected to the quantiles of the bGEV distribution, whereas \(\mu\) and \(\sigma\)
have no simple connections with any kind of moments or quantiles.
Consequently, it is much easier to choose informative priors for \(\mu_\alpha\) and \(\sigma_\beta\).
Based on preliminary experiments, we find that \(\alpha = 0.5\) and \(\beta = 0.8\) are good choices that
makes it easy to select informative priors. This is because the empirical quantiles close to the median
have less variance. We have also experienced that \texttt{R-INLA} is more numerically stable when
the spread is small, i.e. \(\beta\) is large.

\subsection{Models}
\label{sec:model-framework}

Let \(y_t(\bm{s})\) denote the maximum precipitation at location \(\bm{s} \in \mathcal{S}\)
during year \(t \in \mathcal{T}\), where \(\mathcal{S}\) is the study area and \(\mathcal{T}\) is
the time period in focus.
We assume a bGEV distribution for the yearly precipitation maxima,
\begin{linenomath*}
\begin{equation*}
  \label{eq:likelihood}
    \left[y_t(\bm{s}) | \mu_\alpha(\bm{s}), \sigma_\beta(\bm{s}), \xi(\bm{s})\right] \sim
    \text{bGEV}(\mu_\alpha(\bm{s}), \sigma_\beta(\bm{s}), \xi(\bm{s})),
\end{equation*}
\end{linenomath*}
where all observations are assumed to be conditionally independent given the parameters
\(\mu_\alpha(\bm{s})\), \(\sigma_\beta(\bm{s})\) and \(\xi(\bm{s})\). Correct estimation of the tail
parameter is a
difficult problem which highly affects estimates of large quantiles. The tail parameter is assumed
to be constant, i.e.\ \(\xi(\bm{s}) = \xi\). As discussed in Section~\ref{sec:bgev}, this is a common 
procedure, as inference for \(\xi\) is difficult with little data.
The tail parameter is further restricted such that \(\xi < 0.5\), resulting in a finite mean
and variance for the yearly maxima. This restriction makes inference easier and more
numerically stable.
Exploratory analysis of our data supports the hypothesis of a spatially constant \(\xi < 0.5\) and
spatially varying \(\mu_\alpha(\bm{s})\) and \(\sigma_\beta(\bm{s})\) (results not shown).
Two competing models are constructed for describing the spatial structure of \(\mu_\alpha(\bm{s})\)
and \(\sigma_\beta(\bm{s})\).

\subsubsection{The joint model}
\label{sec:joint-model}
In the first model, denoted the joint model, both parameters are modelled
using linear combinations of explanatory variables. Additionally, to draw strength from neighbouring
stations, a spatial Gaussian random field is added to the location parameter. This gives the model
\begin{equation}
  \label{eq:competing-model}
  \begin{aligned}
    \left[y_t(\bm{s}) | \mu_\alpha(\bm{s}), \sigma_\beta(\bm{s}), \xi\right] &\sim
    \text{bGEV}(\mu_\alpha(\bm{s}), \sigma_\beta(\bm{s}), \xi), \\
    \mu_\alpha(\bm{s}) &= \textbf{x}_\mu(\bm{s})^T \bm{\beta}_\mu + u_\mu(\bm{s}), \\
    \log\left(\sigma_\beta(\bm{s})\right) &= \textbf{x}_\sigma(\bm{s})^T \bm{\beta}_\sigma,
  \end{aligned}
\end{equation}
where \(\textbf{x}_\mu(\bm{s})\) and \(\textbf{x}_\sigma(\bm{s})\) are vectors containing an intercept plus
the explanatory variables described in Table~\ref{tab:covariates}, and \(\bm{\beta}_\mu\) and
\(\bm{\beta}_\sigma\) are vectors of regression coefficients. 
The term \(u_\mu(\bm{s})\) is a zero-mean Gaussian field with Matérn correlation function, i.e.,
\begin{linenomath*}
\begin{equation*}
  \label{eq:matern}
    \text{Corr}(u_\mu(\bm{s}_i), u_\mu(\bm{s}_j)) =\frac{1}{2^{\nu - 1} \Gamma(\nu)} \left(\sqrt{8
      \nu} \frac{d(\bm{s}_i, \bm{s}_j)}{\rho}\right)^\nu K_\nu\left(\sqrt{8 \nu} \frac{d(\bm{s}_i,
    \bm{s}_j)}{\rho}\right).
\end{equation*}
\end{linenomath*}
Here, \(d(\bm{s}_i, \bm{s}_j)\) is the Euclidean distance between \(\bm{s}_i\) and \(\bm{s}_j\),
\(\rho > 0\) is the
range parameter and \(\nu > 0\) is the smoothness parameter. The
function \(K_\nu\) is the modified Bessel function of the second kind and order \(\nu\).
The Matérn family is a widely used class of covariance functions in spatial statistics due to its
flexible local behaviour and attractive theoretical properties \citep{stein99_inter,
  matern86_spatial_variat, guttorp06_studies_histor_probab_statis_xlix}. Its form also naturally
appears as the covariance function of some models for the spatial structure of point rain
rates \citep{sun15_mater_model_spatial_covar_struc}.
Efficient inference for high-dimensional Gaussian random fields can be
achieved using the SPDE approach of \citet{lindgren11_explic_link_between_gauss_field}, which is
implemented in \texttt{R-INLA}. It is common to fix the smoothness parameter \(\nu\) instead of
estimating it, as the parameter is difficult to identify from data. The SPDE approximation in
\texttt{R-INLA} allows for \(0 < \nu \leq 1\). We choose \(\nu = 1\) as this 
reflects our beliefs about the smoothness of the underlying physical process. Additionally,
\citet{whittle54_station_proces_plane} argues that \(\nu = 1\) is a more natural choice for
spatial models than the less smooth
exponential correlation function \((\nu = 1/2)\), and \(\nu = 1\) is also the most
extensively tested value when using \texttt{R-INLA} with the SPDE approach
\citep{lindgren15_bayes_spatial_model_with_r_inla}.

The joint model is similar to the models of \citet{geirsson15_comput_effic_spatial_model_annual,
  dyrrdal15_bayes_hierar_model_extrem_hourl_precip_norway,
  davison12_statis_model_spatial_extrem}. However, they all place a Gaussian random field in the
linear predictor for the log-scale and for the tail parameter.
Within the \texttt{R-INLA} framework, it is not possible to model the spread or the tail using Gaussian random fields.
Based on the amount of available data and the difficulty of estimating the spread and tail
parameters, we also believe that the addition of a spatial Gaussian field in either parameter would
simply complicate parameter estimation without any considerable contributions to model performance.
Consequently, we do not include any Gaussian random field in the spread or tail of the bGEV distribution. 

\subsubsection{The two-step model}
\label{sec:twostep-model}

The second model is specifically tailored for sparse data  with large block
sizes. In such data-sparse situations, 
a large observation at a single location can be explained by a large tail parameter or a large spread parameter.
In practice this might cause identifiability issues between \(\sigma_\beta(\bm{s})\)
and \(\xi\), even though the parameters are identifiable in theory. In
order to put a flexible model on the spread while avoiding such issues,
\citet{vandeskog21_model_block_maxim} propose a model which
borrows strength from the peaks over threshold method for separate modelling of
\(\sigma_\beta(\bm{s})\).

For some large enough threshold \(x_{thr}(\bm{s})\), the distribution of sub-daily
precipitation \(X(\bm s)\) larger than \(x_{thr}(\bm{s})\) is assumed to follow a generalised Pareto distribution
\citep{davison90_model_exceed_over_high_thres}
\begin{linenomath*}
\begin{equation*}
  P(X(\bm{s}) > x_{thr}(\bm{s}) + x | X(\bm{s}) > x_{thr}(\bm{s})) = \left(1 + \frac{\xi
      x}{\zeta(\bm{s})}\right)^{-1 / \xi},
\end{equation*}
\end{linenomath*}
with tail parameter \(\xi\) and scale parameter \(\zeta(\bm{s}) = \sigma(\bm{s}) +
\xi(x_{thr}(\bm{s}) - \mu(\bm{s}))\), where \(\mu(\bm s)\) and \(\sigma(\bm s)\) are the original GEV parameters
from \eqref{eq:gev}.
Since \(\xi\) is assumed to be constant in space, all spatial variations in the bGEV distribution
must stem from \(\mu(\bm{s})\) or \(\sigma(\bm{s})\). We therefore assume that the difference
\(x_{thr}(\bm{s}) - \mu(\bm{s})\) between the threshold and the location parameter is proportional to
the scale parameter \(\sigma(\bm{s})\). 
This assumption leads to the spread \(\sigma_\beta(\bm{s})\) being proportional to the standard
deviation of all observations larger than the threshold \(x_{thr}(\bm{s})\).
Based on this assumption, it is possible to model the spatial
structure of the spread parameter independently of the location and tail parameter. Denote
\begin{linenomath*}
\begin{equation*}
  \label{eq:un-standardised-spread}
  \sigma_\beta(\bm{s}) = \sigma_\beta^* \cdot \sigma^*(\bm{s}),
\end{equation*}
\end{linenomath*}
with \(\sigma_\beta^*\) a standardising constant and \(\sigma^*(\bm{s})\) the standard deviation of all
observations larger than \(x_{thr}(\bm{s})\) at location \(\bm{s}\).
Conditional on \(\sigma^*(\bm{s})\), the block maxima can be
standardised as
\begin{linenomath*}
\begin{equation*}
  \label{eq:standardised-response}
  y_t^*(\bm{s}) = y_t(\bm{s}) / \sigma^*(\bm{s}).
\end{equation*}
\end{linenomath*}
The standardised block maxima have a bGEV distribution with a constant spread parameter,
\begin{linenomath*}
\begin{equation*}
  \label{eq:transformed-bgev-likelihood}
  \left[y^*_t(\bm{s}) | \mu_\alpha^*(\bm{s}), \sigma_\beta^*, \xi\right] \sim \text{bGEV}(\mu_\alpha^*(\bm{s}),
  \sigma_\beta^*, \xi),
\end{equation*}
\end{linenomath*}
where \(\mu_\alpha^*(\bm{s}) = \mu_\alpha(\bm{s}) / \sigma^*(\bm{s})\).
Consequently, the second model is divided into two steps. First, we model the standard deviation of large
observations at all locations. Second, we standardise the block maxima observations
and model the remaining parameters of the bGEV distribution. We denote this as the two-step model.
The two-step model shares some similarities with regional frequency analysis
\citep{dalrymple60_flood,hosking97_region,naveau14_fast_nonpar_spatio_tempor_regres,carreau16_charac}, 
which is a multi-step procedure where the data are standardised and pooled together inside
homogeneous regions. However, we standardise the data differently and do not pool together data from
different locations. Instead, we borrow strength from nearby locations by adding a spatial Gaussian
random fields to our model and by keeping \(\xi\) constant for all locations.

The location parameter \(\mu^*_\alpha(\bm{s})\) is modelled as a linear
combination of explanatory variables \(\textbf{x}_\mu(\bm{s})\) and a Gaussian random field
\(u_\mu(\bm{s})\), just as \(\mu_\alpha(\bm{s})\) in the joint model~\eqref{eq:competing-model}.
For estimation of \(\sigma^*(\bm{s})\), the threshold \(x_{thr}(\bm{s})\) is chosen as the \(99\%\) quantile
of all observed precipitation at location \(\bm{s}\). The precipitation observations
larger than \(x_{thr}(\bm{s})\) are declustered to account for temporal dependence, and only the cluster
maximum of an exceedance is used for estimating \(\sigma^*(\bm{s})\).
This might sound counter-intuitive, as the aim of the two-step model is to use more data to simplify
inference. However, even when only using the cluster maxima, inference is less wasteful than for the
joint model.
By using all threshold exceedances for estimating \(\sigma^*(\bm{s})\), we would
need to account for the dependence within exceedance clusters, which would add another layer of complexity to the modelling
procedure. Consequently, we have chosen to not model the temporal dependence and only use the
cluster-maxima for inference in this paper.
To avoid high
uncertainty from locations with few observations, \(\sigma^*(\bm{s})\) is only computed at stations with
more than three years of data. In order to estimate \(\sigma^*(\bm{s})\) at locations with little or no
observations, a linear regression model is used, where the logarithm of \(\sigma^*(\bm{s})\) is
assumed to have a Gaussian distribution,
\begin{linenomath*}
\begin{equation*}
  \label{eq:log-spread-likelihood}
  \left[\log \left(\sigma^*(\bm{s})\right) | \eta(\bm{s}), \tau \right] \sim \mathcal{N}(\eta(\bm{s}),
  \tau^{-1}),
\end{equation*}
\end{linenomath*}
with precision \(\tau\) and mean \(\eta(\bm{s}) = \textbf{x}_\sigma(\bm{s})^T \bm{\beta}_\sigma\).
The estimated posterior mean from the regression model is then used as an estimator for
\(\sigma^*(\bm{s})\) at all locations. Consequently, the complete two-step model is given as
\begin{equation}
  \label{eq:two-step}
  \begin{aligned}
  \left[\log \left(\sigma^*(\bm{s})\right) | \eta(\bm{s}), \tau \right] &\sim \mathcal{N}(\eta(\bm{s}),
  \tau^{-1}), \\
  \eta(\bm{s}) &= \textbf{x}_\sigma(\bm{s})^T \bm{\beta}_\sigma, \\
  \left[y^*_t(\bm{s}) | \mu_\alpha^*(\bm{s}), \sigma_\beta^*, \xi\right] &\sim \text{bGEV}(\mu_\alpha^*(\bm{s}),
  \sigma_\beta^*, \xi), \\
  y^*_t(\bm{s}) &= y_t(\bm{s}) / \sigma^*(\bm{s}), \\
  \mu_\alpha^*(\bm{s}) &= \textbf{x}_\mu(\bm{s})^T \bm{\beta}_\mu + u_\mu(\bm{s}).
  \end{aligned}
\end{equation}

Notice that the formulation of the two-step model makes it trivial to add more complex components
for modelling the spread. One can, therefore, easily add a spatial Gaussian random field to the
linear predictor of \(\log(\sigma^*(\bm{s}))\) while still using the \texttt{R-INLA} framework for
inference, which is not possible with the joint model. 
In Section~\ref{sec:case-study} we perform modelling both with and without a Gaussian random field
in the spread to test how it affects model performance. 

The uncertainty in the estimator for
\(\sigma^*(\bm{s})\) is not propagated into the bGEV model for the standardised
response, meaning that the estimated uncertainties from the two-step model are likely to be too
small. This can be corrected
with a bootstrapping procedure, where we draw \(B\) samples from the posterior
of \(\log (\sigma^*(\bm{s}))\) and estimate \((\mu^*_\alpha(\bm{s}), \sigma^*_\beta, \xi)\) for each
of the \(B\) samples. \citet{vandeskog21_model_block_maxim} show that the two-step model
with 100 bootstrap samples is able to outperform the joint model in a simple setting.

It might seem contradictory to employ a model based on exceedances in our
setting, since we claim that the data quality is too bad to use the peaks over threshold
model for estimating return levels.
However, merely estimating the standard deviation of all threshold exceedances is a much simpler
task than to estimate spatially varying parameters of the generalised Pareto distribution,
including the tail parameter \(\xi\). Thus, while we claim that the available data is not of good
enough quality to estimate return levels in a similar fashion to \citet{opitz18_inla_goes_extrem},
we also claim that it is of good enough quality to perform the simple task of estimating the
trends in the spread parameter. The estimation of all remaining parameters, including \(\xi\), is
performed using block maxima data, which we believe to be of better quality.

\subsection{INLA}
\label{sec:inla}

By placing a Gaussian prior on \(\bm{\beta}_\mu\), both the joint and the
two-step models fall into the class of latent Gaussian models. This is advantageous as it
allows for inference using INLA with the \texttt{R-INLA}
library \citep{rue09_approx_bayes_infer_laten_gauss, rue17_bayes_comput_with_inla,
  bivand2015spatial}. The extreme value framework is quite new to the \texttt{R-INLA}
package. Still, in recent years, some papers have started to appear where it is used for modelling
extremes with INLA \citep[e.g.][]{opitz18_inla_goes_extrem,
  castro-camilo19_splic_gamma_gener_paret_model}. \texttt{R-INLA} includes an implementation of the
SPDE approximation for Gaussian random fields with a Matérn correlation function, which is used on
the random field \(u_\mu(\bm s)\) for a considerable improvement in inference speed.

A requirement for using INLA is that the model likelihood is log-concave. Unfortunately, neither the
GEV distribution nor the bGEV distribution have log-concave likelihoods when \(\xi > 0\). This can
cause severe problems for model inference. However, we find that these problems are mitigated
by choosing slightly informative priors for the model parameters, which is possible because of the
reparametrisation described in Section~\ref{sec:bgev}. Additionally, we find that \texttt{R-INLA} is more
stable when given a response that is standardised such that the difference between its \(95\%\)
quantile and its \(5\%\) quantile is equal to 1. Based on the authors' experience, similar
standardisation of the response is also a common procedure 
when using INLA for estimating the Weibull distribution parameters within the field of survival
analysis. We believe that the combination of slightly informative priors and standardisation of the response is
enough to fix the problems of non-concavity and ensure that \texttt{R-INLA} is working well with the
bGEV distribution.

\subsection{Evaluation}
\label{sec:evaluation}

Model performance can be evaluated using the continuous ranked probability
score \citep[CRPS;][]{matheson76_scorin_rules_contin_probab_distr,
  gneiting07_stric_proper_scorin_rules_predic_estim,friederichs12_forec_verif_extrem_value_distr},
\begin{equation}
  \label{eq:crps}
    \text{CRPS}(F, y) = \int_{-\infty}^{\infty} (F(t) - I(t \geq y))^2 \text{d}t = 2 \int_0^1
    \ell_p\left(y - F^{-1}(p)\right) \text{d}p,
\end{equation}
where \(F\) is the forecast distribution, \(y\) is an observation, \(\ell_p(x) = x (p - I(x < 0))\)
is the quantile loss function and \(I(\cdot)\) is an indicator function.
The CRPS is a strictly proper scoring rule, meaning that the expected value of \(\text{CRPS}(F, y)\) is
minimised for \(G = F\) if and only if \(y \sim G\). The importance of proper
scoring rules when forecasting extremes is discussed by \citet{lerch17_forec_dilem}.
From~\eqref{eq:crps}, one can see that the CRPS is equal to the
integral over the quantile loss function for all possible quantiles. However,
we are only interested in predicting large quantiles, and the model performance for small quantiles
is of little importance to us. The threshold weighted
CRPS \citep[twCRPS;][]{gneiting11_compar_densit_forec_using_thres} is a modification of the CRPS
that allows for emphasis on specific areas of the forecast distribution,
\begin{linenomath*}
\begin{equation}
  \label{eq:twcrps}
  \text{twCRPS}(F, y) = 2 \int_0^1 \ell_p\left(y - F^{-1}(p)\right) w(p) \text{d}p,
\end{equation}
\end{linenomath*}
where \(w(p)\) is a non-negative weight function. A possible choice of \(w(p)\) for focusing
on the right tail is the indicator
function \(w(p) = I(p > p_0)\). As described by \citet{bolin19_scale_depen}, the
mean twCRPS is not robust to outliers and it gives
more weight to forecast distributions with large variances, i.e.\ at locations far away from any
weather station. A scaled version of the twCRPS, denoted the StwCRPS, is created using theorem 5
of \citet{bolin19_scale_depen}:
\begin{linenomath*}
\begin{equation}
  \label{eq:stwcrps-def}
  S_{\text{scaled}}(F, y) = \frac{S(F, y)}{|S(F, F)|} + \log\left(|
    S(F, F)|\right),
\end{equation}
\end{linenomath*}
where \(S(F, y)\) is the twCRPS and \(S(F, F)\) is its expected value with respect to the forecast distribution,
\begin{linenomath*}
\begin{equation*}
  S(F, F) = \int S(F, y) \text{d}F(y).
\end{equation*}
\end{linenomath*}
The mean StwCRPS is more robust to outliers and varying degrees of uncertainty in forecast
distributions, while still being a proper scoring rule \citep{bolin19_scale_depen}.

Using
\texttt{R-INLA} we are able to sample from the posterior distribution of the bGEV parameters at any
location \(\bm{s}\). The forecast distribution at location \(\bm{s}\) is therefore given as
\begin{equation}
  \label{eq:forecast}
  \widehat{F}_{\bm{s}}(\cdot) = \frac{1}{m} \sum_{i = 1}^m F(\cdot; \mu_\alpha^{(i)}(\bm{s}),
  \sigma_\beta^{(i)}(\bm{s}), \xi^{(i)}),
\end{equation}
where \(F\) is the distribution function of the bGEV distribution and \((\mu_\alpha^{(i)}(\bm{s}),
\sigma_\beta^{(i)}(\bm{s}), \xi^{(i)})\) are drawn from the posterior distribution of the bGEV
parameters for \(i = 1, \ldots, m\), where \(m\) is a multiple of the number \(B\) of bootstrap
samples. A closed-form expression is not 
available for the twCRPS when using the forecast distribution from~\eqref{eq:forecast}. Consequently, we
evaluate the twCRPS and StwCRPS using numerical integration.

\section{Modelling sub-daily precipitation extremes in Norway}
\label{sec:case-study}

The models from Section~\ref{sec:method} are applied for estimating return levels in the south of
Norway. Table~\ref{tab:covariates} shows which explanatory variables
are used for modelling the location and spread parameters in both models. All explanatory variables
are standardised to have zero mean and a standard deviation of 1, before being applied for
modelling. Inference for the two-step
model is performed both with and without propagation of the uncertainty in \(\sigma^*(\bm{s})\). The
uncertainty propagation is achieved using 100 bootstrap samples, as described in Section~\ref{sec:twostep-model}. 
Additionally, we modify the two-step model and add a random Gaussian field \(u_\sigma(\bm{s})\) to the
linear predictor of the log-spread, to test if this can yield any considerable improvement in model
performance. Just as \(u_\mu(\bm{s})\), \(u_\sigma(\bm{s})\) has zero mean and a Matérn covariance
function.

\subsection{Prior selection}

Priors must be specified before we can model the precipitation extremes. From construction, the
location parameter \(\mu_\alpha\) is equal to the \(\alpha\) quantile of the
bGEV distribution. This allows us to place a slightly informative prior on \(\bm{\beta}_\mu\), using quantile
regression on \(y^*(\bm{s})\) \citep{koenker05_quant, quantreg}. We choose a Gaussian prior for
\(\bm{\beta}_\mu\), centred at the \(\alpha\) quantile regression estimates and with a precision of
10. There is no unit on the precision in \(\bm{\beta}_\mu\) because the block maxima have been
standardised, as described in Section~\ref{sec:inla}. The regression coefficients
\(\bm{\beta}_\sigma\) differ between the two-step and joint models. In the joint model, all
the coefficients in \(\bm{\beta}_\sigma\), minus the intercept coefficient, are given Gaussian
priors with zero mean and a precision of \(10^{-3}\). The intercept coefficient, here denoted \(\beta_{0, \sigma}\), is
given a log-gamma prior with parameters such that \(\exp(\beta_{0, \sigma})\) has a gamma prior with mean
equal to the empirical difference between the \(1 - \beta/2\) quantile and the \(\beta / 2\)
quantile of the standardised block maxima. The precision of the gamma prior is 10. In the two-step
model, all coefficients of \(\bm{\beta}_\sigma\) are given Gaussian priors with zero mean and a
precision of \(10^{-3}\), while the logarithm of \(\sigma_\beta^*\)
is given the same log-gamma prior as the intercept
coefficient in the joint model.

The parameters of the Gaussian random fields \(u_\mu\) and \(u_\sigma\) are given penalised complexity (PC)
priors. The PC prior is a weakly informative prior distribution, designed to punish model
complexity by placing an exponential prior on the distance from some base
model \citep{simpson17_penal_model_compon_compl}.
\citet{fuglstad19_const_prior_that_penal_compl} develop a joint PC prior for the range \(\rho > 0\)
and standard deviation \(\zeta > 0\) of a Gaussian random field, where the base model is defined to
have infinite range and zero variance. The prior contains two penalty parameters, that can be
decided by specifying the four parameters \(\rho_0\), \(\alpha_1\), \(\zeta_0\) and \(\alpha_2\) such that
\(P(\rho < \rho_0) = \alpha_1\) and \(P(\zeta > \zeta_0) = \alpha_2\).
We choose \(\alpha_1 = \alpha_2 = 0.05\). \(\rho_0\) is given a value of 75 km for both the random
fields, meaning that we place a \(95\%\)
probability on the range being larger than 75 km. To put this range into context, the study area has
a dimension of approximately \(730 \times 460\) km\(^2\), and the mean distance from one station to
its closest neighbour is 10 km. \(\zeta_0\) is given a value of \(0.5\) mm for \(u_\sigma\), meaning
that we place a \(95\%\) probability on the standard deviation being smaller than \(0.5\) mm. This
seems to be a reasonable value because the estimated logarithm of \(\sigma^*(\bm{s})\) lies in the range
between \(0.1\) mm and \(3.5\) mm for all available weather stations and all examined aggregation times. For \(u_\mu\) we
set \(\zeta_0 = 0.5\), which is a reasonable value because of the standardisation of the
response described in Section~\ref{sec:inla}.

A PC prior is also placed on the tail parameter \(\xi\). \citet{opitz18_inla_goes_extrem} develop a
PC prior for the tail parameter of the generalised Pareto distribution, which is the default prior
for \(\xi\) in \texttt{R-INLA} when modelling with the bGEV distribution. However, to the best of our
knowledge, expressions for the PC priors for \(\xi\) in the GEV or bGEV distributions are not
previously available in the literature. In the supplementary material we develop expressions for the PC prior
of \(\xi \in [0, 1)\) with base model \(\xi = 0\) for the GEV distribution and the bGEV
distribution. Closed-form expressions do not exist, but the priors can be approximated
numerically. Having computed the PC priors for the GEV distribution and the 
bGEV distribution, we find that they are similar to the PC prior of the generalised Pareto
distribution, which has a closed-form expression and is already implemented in
\texttt{R-INLA}. Consequently, we choose to model the tail parameter of the bGEV distribution
with the PC prior for the generalised Pareto distribution \citep{opitz18_inla_goes_extrem}:
\begin{linenomath*}
\begin{equation*}
  \label{eq:gp-pc-prior}
  \pi(\xi) = \frac{\lambda}{\sqrt{2}} \exp\left(-\frac{\lambda}{\sqrt{2}}\frac{\xi}{(1 -
      \xi)^{1/2}}\right) \left(\frac{1 - \xi/2}{(1 - \xi)^{3/2}}\right),
\end{equation*}
\end{linenomath*}
with \(0 \leq \xi < 1\) and penalty parameter \(\lambda\). Even though the prior is defined for
values of \(\xi\) up to 1, a reparametrisation is performed within \texttt{R-INLA} such that \(0 \leq
\xi < 0.5\).
Since the base model has \(\xi = 0\), the prior places more weight on small values of \(\xi\)
when \(\lambda\) increases. Based on the plots in Figure~S2.1 in the supplementary material, we find
a value of \(\lambda = 7\) to give a good description of our prior beliefs, as we expect \(\xi\) to
be positive but small.

\subsection{Cross-validation}

Model performance is evaluated using five-fold cross-validation with the StwCRPS. The StwCRPS weight
function is chosen as \(w(p) = I(p > 0.9)\). Both in-sample and 
out-of-sample performance are evaluated. The mean StwCRPS over all five
folds are displayed in Table~\ref{tab:twcrps}. The two-step model outperforms the
joint model for all aggregation times. This implies
that information about threshold exceedances can provide valuable information when modelling block maxima.
When performing in-sample estimation, the variant of the
two-step model with a Gaussian field and without bootstrapping always outperforms the other
contestants.
However, during out-of-sample estimation, the model performs worse than its competitors.
This indicates a tendency to overfit when not using bootstrapping to
propagate uncertainty in \(\sigma^*(\bm{s})\) into the estimation of \((\mu_\alpha^*(\bm{s}),
\sigma^*_\beta, \xi)\). 
The two variants of the two-step model that use bootstrapping perform best during
out-of-sample estimation.
While their model fits yield similar scores, their difference in complexity is quite considerable,
as one model contains two spatial random fields, and the other only contains one. This shows that
there is little need for placing an overly complex model on the spread parameter.
Consequently, for estimation of the bGEV parameters and return levels, we choose to use the
two-step model with bootstrapping and without a spatial Gaussian random field in the spread.

\begin{table}
  \centering
  \renewcommand*{\arraystretch}{1.4}
  \caption{Mean StwCRPS with weight function \(w(p) = I(p > 0.9)\) for 5-fold cross-validation
    performed using out-of-sample estimation and in-sample estimation. The two-step method is tested with and without a Gaussian field in the spread and bootstrapping for propagation of uncertainty.
    Cross-validation is performed
    for precipitation aggregated over periods of 1 hour, 3 hours, 6 hours, 12 hours and 24
    hours. The best scores are written in \textbf{bold}.}
  \label{tab:twcrps}
  \begin{tabular}{llccrrrrr}
    \toprule
    & Model & \(u_\sigma(\bm{s})\) & Boot- & 1 h & 3 h & 6 h & 12 h & 24 h \\[-.5em]
    & & & strap & & & & & \\
    \midrule
Out-of-& Joint & & & \(-0.872\) & \(-0.597\) & \(-0.412\) & \(-0.149\) & \( 0.0500\) \\
  sample & Two-step & \checkmark & \checkmark & \(\bm{-0.905}\) & \(\bm{-0.603}\) & \(-0.427\) & \(-0.207\) & \(\bm{ 0.0425}\) \\
  & Two-step & \checkmark & & \(-0.893\) & \(-0.585\) & \(-0.414\) & \(-0.197\) & \( 0.0635\) \\
  & Two-step & & \checkmark & \(-0.876\) & \(-0.594\) & \(\bm{-0.429}\) & \(\bm{-0.211}\) & \( 0.0456\) \\
  & Two-step & &  & \(-0.872\) & \(-0.584\) & \(-0.417\) & \(-0.196\) & \( 0.0674\) \\
 \midrule
In- & Joint & & & \(-0.876\) & \(-0.608\) & \(-0.445\) & \(-0.230\) & \( 0.0206\) \\
  sample & Two-step & \checkmark & \checkmark & \(-1.004\) & \(-0.713\) & \(-0.564\) & \(-0.328\) & \(-0.1066\) \\
  & Two-step & \checkmark & & \(\bm{-1.012}\) & \(\bm{-0.721}\) & \(\bm{-0.577}\) & \(\bm{-0.333}\) & \(\bm{-0.1161}\) \\
  & Two-step & & \checkmark & \(-0.889\) & \(-0.607\) & \(-0.453\) & \(-0.247\) & \(-0.0244\) \\
  & Two-step & &  & \(-0.886\) & \(-0.607\) & \(-0.454\) & \(-0.243\) & \(-0.0182\) \\
  \end{tabular}
\end{table}

\subsection{Parameter estimates}

The parameters of the two-step model are estimated for different aggregation times between 1 and 24 hours.
Uncertainty is propagated using \(B = 100\) bootstrap samples. Estimation of the posterior of
\((\mu_\alpha^*(\bm{s}), \sigma^*_\beta, \xi)\) given some value of \(\sigma^*(\bm s)\) takes less than 2
minutes on a 2.4 gHz laptop with 16 GB RAM, and
the 100 bootstraps can be computed in parallel. On a moderately sized computational server,
inference can thus be performed in well under 10 minutes.

The estimated values of the regression coefficients \(\bm{\beta}_\mu\) and \(\bm{\beta}_\sigma\),
the spread \(\sigma_\beta^*\) and the standard deviation of the Gaussian field
\(u_\mu(\bm{s})\) for the standardised precipitation maxima, are displayed in
Table~\ref{tab:beta_vals} for some selected temporal aggregations. These estimates are computed
by drawing 20 samples from each of the \(100\) posterior distributions. The empirical mean, standard
deviation and quantiles of these 2000 samples are then reported.
\begin{table}
  \centering
  \renewcommand*{\arraystretch}{1.2}
  \caption{Estimated regression coefficients \(\bm{\beta}_\mu\), \(\bm{\beta}_\sigma\) and estimated
    standard deviation \(\text{SD}(u_\mu)\) of the Gaussian field \(u_\mu(\bm{s})\)
    in the two-step model for yearly maximum precipitation at different temporal
    aggregations.}
  \label{tab:beta_vals}
  \resizebox{\textwidth}{!}{%
    \begin{tabular}{lllrrrrr}
      \toprule 
      Temporal & Parameter & Explanatory & Mean & SD & \(2.5\%\) & \(50\%\) & \(97.5\%\) \\[-.3em]
      aggregation & & variable & & & quantile & quantile & quantile \\
      \midrule
      1 hour & \(\bm{\beta}_\mu\) & Intercept & \( 0.669\) & \(0.066\) & \( 0.518\) & \( 0.675\) & \(0.794\) \\
               & & Mean annual precipitation & \( 0.071\) & \(0.012\) & \( 0.044\) & \( 0.072\) & \(0.091\) \\
               & & Altitude & \(-0.006\) & \(0.011\) & \(-0.026\) & \(-0.006\) & \(0.013\) \\
               & & Easting & \(-0.044\) & \(0.035\) & \(-0.118\) & \(-0.042\) & \(0.017\) \\
               & & Northing & \( 0.036\) & \(0.038\) & \(-0.036\) & \( 0.036\) & \(0.111\) \\
               & & Distance to the open sea & \(-0.001\) & \(0.033\) & \(-0.072\) & \( 0.002\) & \(0.055\) \\
               & \(\text{SD}(u_\mu)\) & & \(0.084\) & \(0.023\) & \(0.047\) & \(0.085\) & \(0.127\) \\
               & \(\sigma^*_\beta\) & & \(0.118\) & \(0.002\) & \(0.113\) & \(0.118\) & \(0.123\) \\
               & \(\bm{\beta}_\sigma\) & Intercept & \( 2.836\) & \(0.007\) & \( 2.822\) & \( 2.831\) & \( 2.835\) \\
               & & Easting & \( 0.088\) & \(0.015\) & \( 0.058\) & \( 0.078\) & \( 0.087\) \\
               & & Northing & \(-0.150\) & \(0.015\) & \(-0.177\) & \(-0.160\) & \(-0.151\) \\
               & & Distance to the open sea & \(-0.050\) & \(0.017\) & \(-0.080\) & \(-0.062\) & \(-0.051\) \\
      \midrule
      3 hours & \(\bm{\beta}_\mu\) & Intercept & \( 0.847\) & \(0.025\) & \( 0.800\) & \( 0.846\) & \(0.895\) \\
               & & Mean annual precipitation & \( 0.120\) & \(0.012\) & \( 0.096\) & \( 0.122\) & \(0.139\) \\
               & & Altitude & \(-0.009\) & \(0.009\) & \(-0.027\) & \(-0.010\) & \(0.007\) \\
               & & Easting & \( 0.016\) & \(0.021\) & \(-0.025\) & \( 0.017\) & \(0.057\) \\
               & & Northing & \( 0.022\) & \(0.018\) & \(-0.015\) & \( 0.024\) & \(0.054\) \\
               & & Distance to the open sea & \( 0.017\) & \(0.018\) & \(-0.021\) & \( 0.017\) & \(0.052\) \\
               & \(\text{SD}(u_\mu)\) & & \(0.062\) & \(0.013\) & \(0.045\) & \(0.060\) & \(0.085\) \\
               & \(\sigma^*_\beta\) & & \(0.123\) & \(0.003\) & \(0.117\) & \(0.123\) & \(0.129\) \\
               & \(\bm{\beta}_\sigma\) & Intercept & \( 3.207\) & \(0.017\) & \( 3.177\) & \( 3.190\) & \( 3.208\) \\
               & & Easting & \( 0.031\) & \(0.014\) & \( 0.003\) & \( 0.021\) & \( 0.030\) \\
               & & Northing & \(-0.138\) & \(0.014\) & \(-0.163\) & \(-0.147\) & \(-0.138\) \\
               & & Distance to the open sea & \(-0.057\) & \(0.016\) & \(-0.083\) & \(-0.068\) & \(-0.058\) \\
      \midrule
      6 hours & \(\bm{\beta}_\mu\) & Intercept & \( 0.911\) & \(0.024\) & \( 0.868\) & \( 0.911\) & \(0.962\) \\
               & & Mean annual precipitation & \( 0.148\) & \(0.013\) & \( 0.123\) & \( 0.150\) & \(0.168\) \\
               & & Altitude & \(-0.012\) & \(0.009\) & \(-0.029\) & \(-0.011\) & \(0.005\) \\
               & & Easting & \( 0.040\) & \(0.021\) & \(-0.001\) & \( 0.040\) & \(0.080\) \\
               & & Northing & \( 0.008\) & \(0.019\) & \(-0.029\) & \( 0.009\) & \(0.042\) \\
               & & Distance to the open sea & \( 0.030\) & \(0.020\) & \(-0.011\) & \( 0.031\) & \(0.067\) \\
               & \(\text{SD}(u_\mu)\) & & \(0.067\) & \(0.011\) & \(0.051\) & \(0.065\) & \(0.090\) \\
               & \(\sigma^*_\beta\) & & \(0.125\) & \(0.004\) & \(0.118\) & \(0.125\) & \(0.133\) \\
               & \(\bm{\beta}_\sigma\) & Intercept & \( 3.470\) & \(0.020\) & \( 3.435\) & \( 3.456\) & \( 3.469\) \\
               & & Easting & \(-0.032\) & \(0.015\) & \(-0.062\) & \(-0.042\) & \(-0.033\) \\
               & & Northing & \(-0.095\) & \(0.014\) & \(-0.122\) & \(-0.105\) & \(-0.096\) \\
               & & Distance to the open sea & \(-0.065\) & \(0.016\) & \(-0.093\) & \(-0.077\) & \(-0.066\) \\
      \bottomrule
  \end{tabular}
  }
\end{table}
There is strong evidence that all the explanatory variables in \(\textbf{x}_\sigma(\bm{s})\) are affecting the spread, with the northing being the most important explanatory variable.
There is considerably less evidence that all our chosen explanatory variables have an effect on the location parameter.
However, as the posterior distribution of \(\bm{\beta}_\mu\) is estimated using 100 different samples
from the posterior of \(\sigma^*(\bm{s})\), it might be that the different regression coefficients
are more significant for some of the standardisations, and less significant for others. 
The explanatory variable that has the greatest effect on the location parameter seems to be the mean annual precipitation. 
Thus, at locations with large amounts of precipitation, we expect the extreme precipitation to be
heavier than at locations with little precipitation. From the estimates for \(\bm{\beta}_\sigma\), we
also expect more variance in the distribution of extreme precipitation in the south. 
The standard deviation of \(u_\mu(\bm{s})\) is of approximately the same magnitude as most of
the regression coefficients in \(\bm{\beta}_\mu\).
\begin{table}
  \centering
  \renewcommand*{\arraystretch}{1.4}
  \caption{Estimated posterior mean and quantiles for the range \(\rho\) of the Gaussian field
    \(u_\mu(\bm{s})\) and the tail parameter \(\xi\) in the two-step model for yearly maximum
    precipitation at different temporal aggregations.}
  \label{tab:range-xi-posteriors}
  \begin{tabular}{llllll}\toprule
    Parameter & Temporal & Mean & \(2.5\%\) & \(50\%\) &  \(97.5\%\) \\[-.5em]
              & aggregation &  & quantile & quantile & quantile \\
    \midrule
    \(\rho\) & 1 hour & \(235\) & \(34\) & \(255\) & \(478\) \\
    & 3 hours & \(78\) & \(39\) & \(75\) & \(147\) \\
    & 6 hours & \(60\) & \(32\) & \(57\) & \(104\) \\
    & 12 hours & \(84\) & \(31\) & \(83\) & \(145\) \\
    & 24 hours & \(55\) & \(32\) & \(50\) & \(105\) \\
    \midrule
    \(\xi\) & 1 hour & \(0.178\) & \(0.136\) & \(0.179\) & \(0.211\) \\
    & 3 hours & \(0.090\) & \(0.057\) & \(0.089\) & \(0.120\) \\
    & 6 hours & \(0.047\) & \(0.028\) & \(0.046\) & \(0.072\) \\
    & 12 hours & \(0.032\) & \(0.010\) & \(0.031\) & \(0.048\) \\
    & 24 hours & \(0.029\) & \(0.006\) & \(0.029\) & \(0.051\) \\
    \bottomrule
  \end{tabular}
\end{table}

\begin{figure}
  \centering
  \includegraphics[width=.9\linewidth]{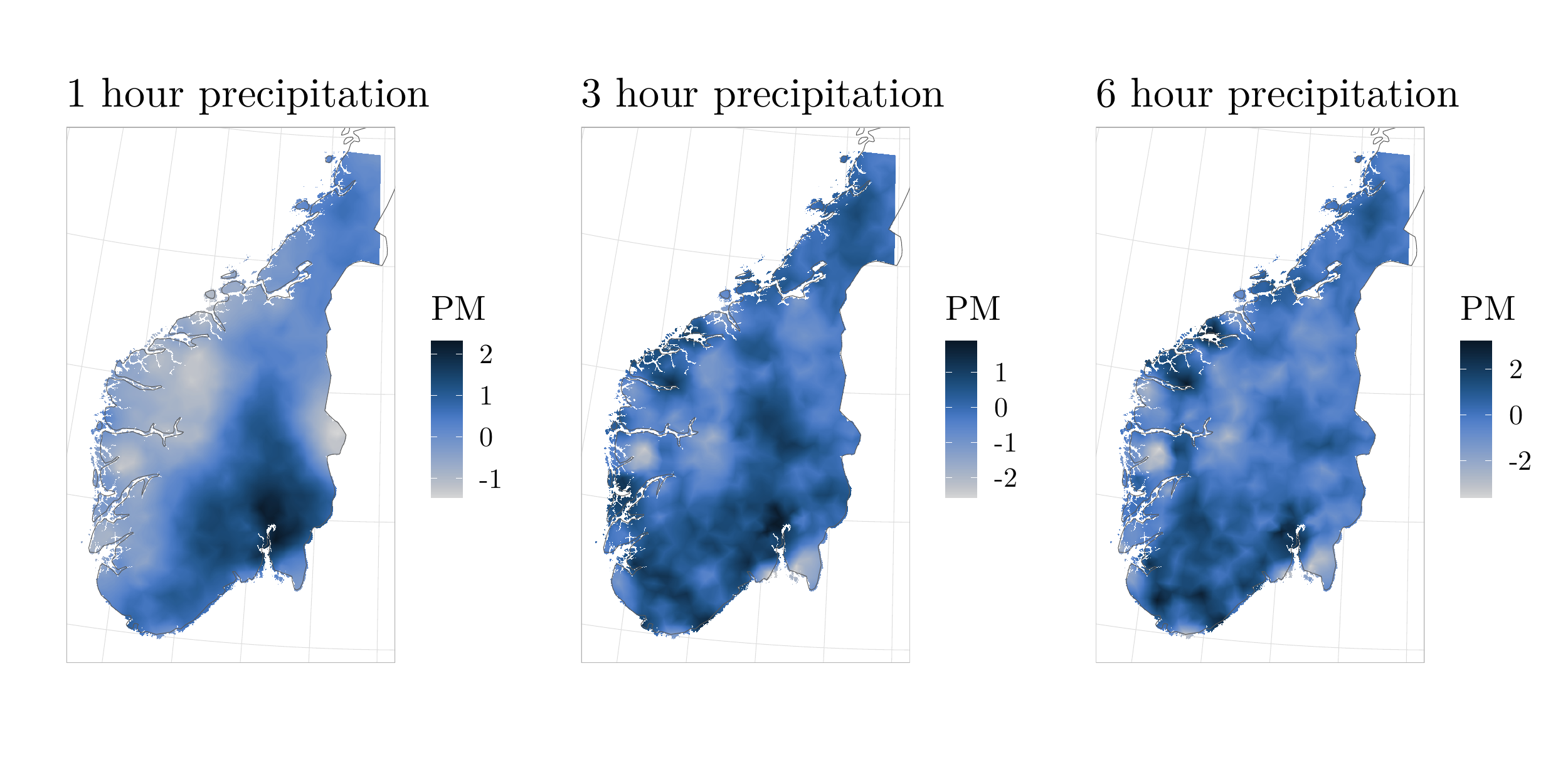}
  \caption{Estimated posterior mean (PM) of the Gaussian field \(u_\mu(\bm{s})\) for three different
    temporal aggregations of precipitation, with unit mm.}
  \label{fig:matern-plot}
\end{figure}

Table~\ref{tab:range-xi-posteriors} displays the posterior range of the \(u_\mu(\bm{s})\). For the
available data, the median number of neighbours within a radius of 50 km is 17, and the
median number of neighbours within a radius of 100 km is 36. Based on these numbers, one can see
that the Gaussian field is able to introduce spatial correlation between a large number of different
stations.
The range of the Gaussian field is considerably reduced as the temporal aggregation increases.
It seems that, for 1 hour precipitation, the regression coefficients are unable to explain some kind
of large-scale phenomenon that considerably affects the location parameter \(\mu_\alpha(\bm{s})\). To
correct this, the range of \(u_\mu(\bm{s})\) has to be large.
For longer aggregation periods, this phenomenon is not as important anymore, and the regression
coefficients are able to explain most of the large-scale trends. Consequently, the range of
\(u_\mu(\bm{s})\) is decreased.
The posterior means of \(u_\mu(\bm{s})\) for three different temporal aggregations are displayed over
a \(1\times1\) km\(^2\) gridded map in Figure~\ref{fig:matern-plot}.
It is known that extreme precipitation dominates in the
southeast of Norway for short aggregation times because of its large amount of convective
precipitation (see, e.g, \cite{dyrrdal15_bayes_hierar_model_extrem_hourl_precip_norway}). Based on
Figure~\ref{fig:matern-plot} it becomes evident that our explanatory variables are unable to
describe this regional difference when modelling hourly precipitation, and \(u_\mu(\bm{s})\) has
to do the job of separating between the east and the west.
As the temporal aggregations
increase from one hour to three and six hours, the difference between east and west diminishes,
and it seems that the explanatory variables do a better job of explaining the trends in the
location parameter \(\mu_\alpha(\bm s)\).

The
posterior distribution of \(\xi\) is also described in Table~\ref{tab:range-xi-posteriors}. The tail parameter
seems to decrease quickly as the aggregation time increases, and it is practically constant for precipitation over longer
periods than 12 hours. This makes sense given the observation of
\citet{barbero19_synth_hourl_daily_precip_extrem} that most 24 hour annual maximum precipitation
comes from rainstorms with lengths of less than 15 hours. Thus, the tail parameter for 24 hour precipitation should be close to the tail parameter for 12 hour precipitation. For
12 hours and up, the tail parameter is so small that one may 
wonder if a Gumbel distribution would not have given a better fit to the data.
However, this is not the case for the shorter aggregation times, where the tail parameter is considerably larger.

\subsection{Return levels}

\begin{figure*}
  \centering
  \includegraphics[width=.9\linewidth]{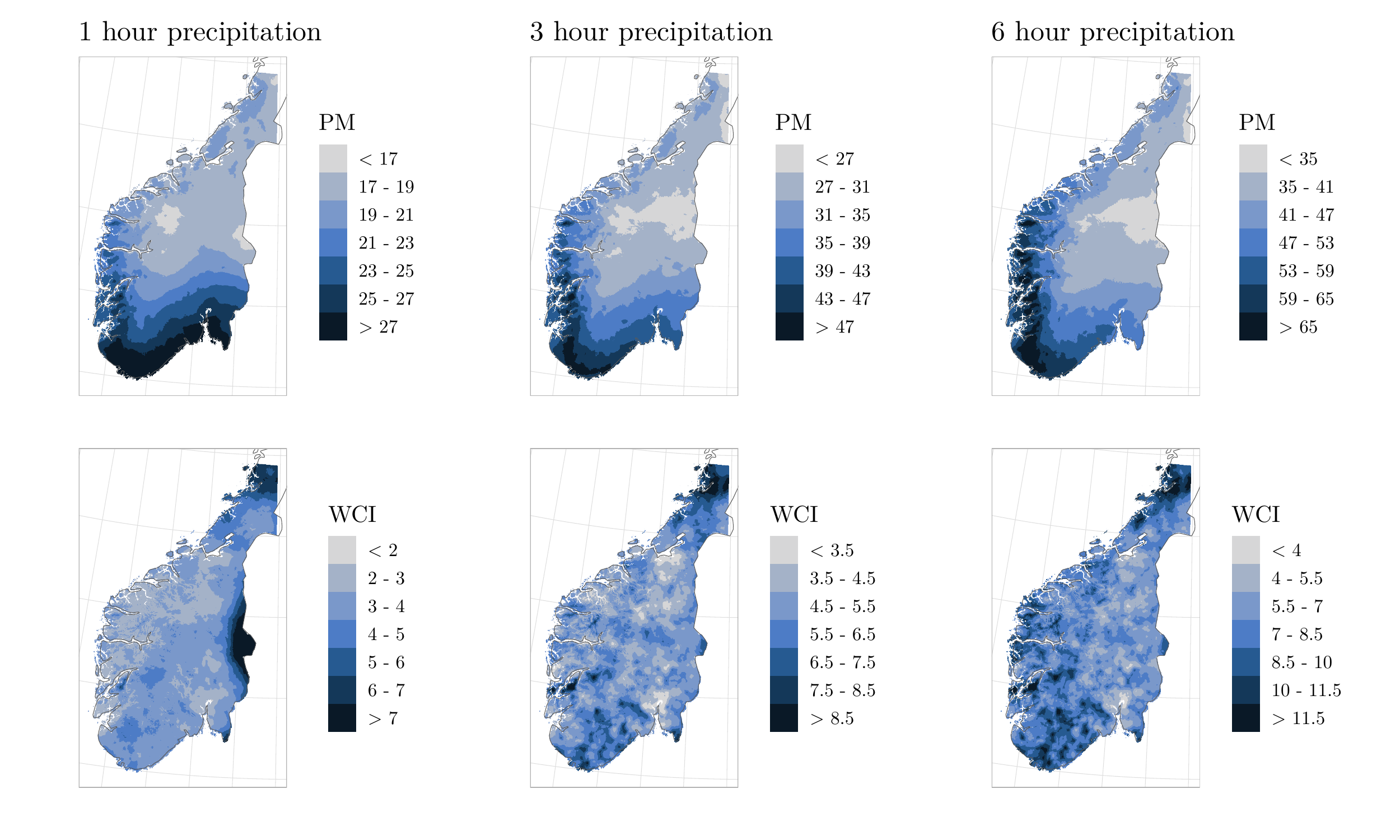}
  \caption{Estimated posterior mean (PM) and width of the \(95\%\) credible intervals (WCI) for the
    20 year return levels of sub-daily precipitation. Different aggregation times are displayed in
    different columns. All numbers are given with unit mm.} 
  \label{fig:return-levels}
\end{figure*}

We use the two-step model for estimating large return levels for the yearly precipitation
maxima. Posterior distributions of the 20 year return levels
are estimated on a grid with resolution \(1\times 1\) km\(^2\). The posterior
means and the widths of the \(95\%\) credible intervals are displayed in
Figure~\ref{fig:return-levels}. For a period of 1 hour the most extreme precipitation is located
southeast in Norway, while for longer periods, the extreme precipitation is moving over to the west
coast. These results are expected since we know that the convective precipitation of the southeast dominates for short aggregation periods. At the same time, the southwest of Norway generally has more precipitation, making it the
dominant region for longer aggregation
times. The spatial structure
and magnitude of the 20 year return levels for hourly precipitation are similar to the estimates
of \citet{dyrrdal15_bayes_hierar_model_extrem_hourl_precip_norway}, but with considerably thinner
credible intervals. This makes sense as more data are available, and the two-step model is able to perform less wasteful inference. In addition, our model is much more simple, as they include a random Gaussian field in all three parameters, while we only include a random Gaussian field in the location parameter. This can also lead to less uncertainty in the return level estimates.

\section{Conclusion}
\label{sec:conclusion}

The blended generalised extreme value (bGEV) distribution is applied as a substitute for the
generalised extreme value (GEV) distribution for estimation of the return levels of sub-daily
precipitation in the south of Norway. The bGEV distribution simplifies inference
by introducing a parameter-free support, but can only be applied for modelling of heavy-tailed
phenomena. Sub-daily precipitation maxima are modelled using a spatial Bayesian hierarchical model
with a latent Gaussian field. This is implemented using
both integrated nested Laplace approximations (INLA) and the stochastic partial differential
equation (SPDE) approach, for fast inference.
Inference is also made more stable and less wasteful by our novel two-step modelling procedure that
borrows strength from the peaks over threshold method when modelling block maxima.
Like the GEV distribution,
the bGEV distribution suffers from a lack of log-concavity, which can cause problems when using
INLA. We are able to mitigate any problems caused by a lack of log-concavity by choosing
slightly informative priors and standardising the data.
We find that the bGEV distribution performs well as a model for extreme precipitation. The two-step
model
successfully utilises the additional information provided by the peaks over threshold data and is
able to outperform models that only use block maxima data for inference.

\section*{Acknowledgements}
We thank Thordis L. Thorarinsdottir and Geir-Arne Fuglstad for helpful discussions.

\section*{Funding}
This publication is part of the World of Wild Waters (WoWW) project, which falls under the umbrella of Norwegian University of Science and Technology (NTNU)’s Digital Transformation initiative.

\section*{Code and data availability}
The necessary code and data for achieving these results are available online at \url{https://github.com/siliusmv/inlaBGEV}. The unprocessed data are freely available online, as described in Section~\ref{sec:data}.

\section*{Conflicts of interest}
The authors declare that they have no conflicts of interest.

\renewcommand*{\bibfont}{\footnotesize}

\printbibliography

@article{bivand2015spatial,
  title={Spatial Data Analysis with {R-INLA} with Some Extensions},
  author={Bivand, Roger and G{\'o}mez-Rubio, Virgilio and Rue, H{\aa}vard},
  journal={Journal of Statistical Software},
  volume={63},
  number={1},
  pages={1--31},
  year={2015},
  publisher={Directory of Open Access Journals}
}

@article{buecher21_horse_race_between_block_maxim,
  author =	 {Axel B{\"u}cher and Chen Zhou},
  title =	 {{A Horse Race Between the Block Maxima Method and the Peak-Over-Threshold
                  Approach}},
  journal =	 {Statistical Science},
  volume =	 36,
  number =	 3,
  pages =	 {360 -- 378},
  year =	 2021,
  doi =		 {10.1214/20-STS795},
  url =		 {https://doi.org/10.1214/20-STS795},
  publisher =	 {Institute of Mathematical Statistics},
}

@Article{lussana18_daily_precip_obser_gridd_datas,
  author    = {Lussana, C and Saloranta, Tuomo and Skaugen, Thomas and Magnusson, Jan and Tveito, Ole Einar and Andersen, Jess},
  title     = {se{N}orge2 Daily Precipitation, an Observational Gridded Dataset Over {N}orway From 1957 To the Present Day},
  doi       = {10.5194/essd-10-235-2018},
  number    = {1},
  pages     = {235},
  url       = {https://doi.org/10.5194/essd-10-235-2018},
  volume    = {10},
  journal   = {Earth Syst. Sci. Data},
  publisher = {Copernicus GmbH},
  year      = {2018},
}

@Article{lussana18_three_dimen_spatial_inter_m,
  author    = {Lussana, C and Tveito, OE and Uboldi, F},
  title     = {Three-Dimensional Spatial Interpolation of 2 M Temperature Over {N}orway},
  doi       = {10.1002/qj.3208},
  number    = {711},
  pages     = {344--364},
  url       = {https://doi.org/10.1002/qj.3208},
  volume    = {144},
  journal   = {Q. J. R. Meteorolog. Soc.},
  publisher = {Wiley Online Library},
  year      = {2018},
}

@book{koenker05_quant,
  author =	 {Koenker, Roger},
  title =	 {Quantile regression},
  year =	 {2005},
  url =		 {https://doi.org/10.1017/CBO9780511754098},
  address =	 {Cambridge},
  doi =		 {10.1017/CBO9780511754098},
  isbn =	 {9780511754098},
  series =	 {Econometric Society monographs ;},
  volume =	 {38},
}

@article{smith85_maxim_likel_estim_class_nonreg,
  author =	 {Smith, Richard L.},
  title =	 {Maximum Likelihood Estimation in a Class of Nonregular
                  cases},
  journal =	 {Biometrika},
  volume =	 72,
  number =	 1,
  pages =	 {67--90},
  year =	 1985,
  doi =		 {10.1093/biomet/72.1.67},
  url =		 {https://doi.org/10.1093/biomet/72.1.67},
}

@book{coles01_introd_statis_model_extrem_values,
  author =	 {Coles, Stuart},
  title =	 {An Introduction to Statistical Modeling of Extreme
                  Values},
  year =	 2001,
  publisher =	 {Springer, London},
  url =		 {https://doi.org/10.1007/978-1-4471-3675-0},
  doi =		 {10.1007/978-1-4471-3675-0},
  isbn =	 {978-1-84996-874-4},
}

@Article{davison15_statis_extrem,
  author  = {Davison, A. C. and Huser, R.},
  title   = {Statistics of Extremes},
  doi     = {10.1146/annurev-statistics-010814-020133},
  number  = {1},
  pages   = {203-235},
  url     = {https://doi.org/10.1146/annurev-statistics-010814-020133},
  volume  = {2},
  journal = {Annu. Rev. Stat. Appl.},
  year    = {2015},
}

@article{davison12_statis_model_spatial_extrem,
  author =	 "Davison, A. C. and Padoan, S. A. and Ribatet, M.",
  title =	 {Statistical Modeling of Spatial Extremes},
  journal =	 "Statist. Sci.",
  volume =	 27,
  number =	 2,
  pages =	 "161--186",
  year =	 2012,
  doi =		 "10.1214/11-STS376",
  url =		 {https://doi.org/10.1214/11-STS376},
  fjournal =	 "Statistical Science",
  month =	 05,
  publisher =	 "The Institute of Mathematical Statistics",
}

@Article{rue09_approx_bayes_infer_laten_gauss,
  author  = {Rue, H{\aa}vard and Martino, Sara and Chopin, Nicolas},
  title   = {Approximate {B}ayesian Inference for Latent {G}aussian Models By Using Integrated Nested {L}aplace Approximations},
  doi     = {10.1111/j.1467-9868.2008.00700.x},
  number  = {2},
  pages   = {319-392},
  url     = {https://doi.org/10.1111/j.1467-9868.2008.00700.x},
  volume  = {71},
  journal = {J. R. Stat. Soc. B},
  year    = {2009},
}

@Article{opitz18_inla_goes_extrem,
  author =	 "Opitz, Thomas and Huser, Rapha{\"e}l and Bakka, Haakon
                  and Rue, H{\aa}vard",
  title =	 {{INLA} Goes Extreme: {B}ayesian Tail Regression for the
                  Estimation of High Spatio-Temporal Quantiles},
  journal =	 "Extremes",
  volume =	 21,
  number =	 3,
  pages =	 "441--462",
  year =	 2018,
  doi =		 "10.1007/s10687-018-0324-x",
  url =		 {https://doi.org/10.1007/s10687-018-0324-x},
  issn =	 "1572-915X",
}

@article{dyrrdal15_bayes_hierar_model_extrem_hourl_precip_norway,
  author =	 {Dyrrdal, Anita Verpe and Lenkoski, Alex and
                  Thorarinsdottir, Thordis L. and Stordal, Frode},
  title =	 {{B}ayesian Hierarchical Modeling of Extreme Hourly
                  Precipitation in {N}orway},
  journal =	 {Environmetrics},
  volume =	 26,
  number =	 2,
  pages =	 {89-106},
  year =	 2015,
  doi =		 {10.1002/env.2301},
  url =		 {https://doi.org/10.1002/env.2301},
}

@Article{davison90_model_exceed_over_high_thres,
  author  = {Davison, A. C. and Smith, R. L.},
  title   = {Models for Exceedances Over High Thresholds},
  doi     = {10.1111/j.2517-6161.1990.tb01796.x},
  number  = {3},
  pages   = {393-425},
  url     = {https://doi.org/10.1111/j.2517-6161.1990.tb01796.x},
  volume  = {52},
  journal = {J. R. Stat. Soc. B},
  year    = {1990},
}

@Article{fisher28_limit_forms_frequen_distr_larges,
  author    = {Fisher, R. A. and Tippett, L. H. C.},
  title     = {Limiting Forms of the Frequency Distribution of the Largest Or Smallest Member of a Sample},
  doi       = {10.1017/S0305004100015681},
  number    = {2},
  pages     = {180-190},
  volume    = {24},
  journal   = {Math. Proc. Cambridge Philos. Soc.},
  publisher = {Cambridge University Press},
  year      = {1928},
}

@article{crespi18_high_resol_month_precip_climat_over_norway,
  author =	 {Alice Crespi and Cristian Lussana and Michele Brunetti and Andreas Dobler and
                  Maurizio Maugeri and Ole Einar Tveito},
  title =	 {High-Resolution Monthly Precipitation Climatologies Over {N}orway: Assessment of
                  Spatial Interpolation Methods},
  journal =	 {arXiv:\allowbreak 1804.04867},
  year =	 2018,
}

@Article{lindgren11_explic_link_between_gauss_field,
  author  = {Lindgren, Finn and Rue, H{\aa}vard and Lindstr{\"o}m, Johan},
  title   = {An Explicit Link Between {G}aussian Fields and {G}aussian {M}arkov Random Fields: the Stochastic Partial Differential Equation Approach},
  doi     = {10.1111/j.1467-9868.2011.00777.x},
  number  = {4},
  pages   = {423-498},
  url     = {https://doi.org/10.1111/j.1467-9868.2011.00777.x},
  volume  = {73},
  journal = {J. R. Stat. Soc. B},
  year    = {2011},
}

@Article{lindgren15_bayes_spatial_model_with_r_inla,
  author    = {Finn Lindgren and H{\aa}vard Rue},
  title     = {{B}ayesian Spatial Modelling With {R-INLA}},
  issn      = {1548-7660},
  language  = {English},
  number    = {19},
  volume    = {63},
  journal   = {J Stat Softw},
  month     = {2},
  publisher = {University of California at Los Angeles},
  year      = {2015},
}

@Article{rue17_bayes_comput_with_inla,
  author  = {Rue, H{\aa}vard and Riebler, Andrea and S{\o}rbye, Sigrunn H. and Illian, Janine B. and Simpson, Daniel P. and Lindgren, Finn K.},
  title   = {{B}ayesian Computing With {INLA}: a Review},
  doi     = {10.1146/annurev-statistics-060116-054045},
  number  = {1},
  pages   = {395-421},
  url     = {https://doi.org/10.1146/annurev-statistics-060116-054045},
  volume  = {4},
  journal = {Annu. Rev. Stat. Appl.},
  year    = {2017},
}

@Article{bakka18_spatial_model_with_r_inla,
  author  = {Bakka, Haakon and Rue, H{\aa}vard and Fuglstad, Geir-Arne and Riebler, Andrea and Bolin, David and Illian, Janine and Krainski, Elias and Simpson, Daniel and Lindgren, Finn},
  title   = {Spatial Modeling With {R-INLA}: a Review},
  doi     = {10.1002/wics.1443},
  number  = {6},
  pages   = {e1443},
  url     = {https://doi.org/10.1002/wics.1443},
  volume  = {10},
  journal = {WIREs Comput. Stat.},
  year    = {2018},
}

@article{simpson17_penal_model_compon_compl,
  author =	 "Simpson, Daniel and Rue, H{\aa}vard and Riebler, Andrea
                  and Martins, Thiago G. and S{\o}rbye, Sigrunn H.",
  title =	 {Penalising Model Component Complexity: a Principled,
                  Practical Approach To Constructing Priors},
  journal =	 "Statist. Sci.",
  volume =	 32,
  number =	 1,
  pages =	 "1--28",
  year =	 2017,
  doi =		 "10.1214/16-STS576",
  url =		 {https://doi.org/10.1214/16-STS576},
  fjournal =	 "Statistical Science",
  month =	 02,
  publisher =	 "The Institute of Mathematical Statistics",
}

@Article{fuglstad19_const_prior_that_penal_compl,
  author    = {Geir-Arne Fuglstad and Daniel Simpson and Finn Lindgren and H{\aa}vard Rue},
  title     = {Constructing Priors That Penalize the Complexity of {G}aussian Random Fields},
  doi       = {10.1080/01621459.2017.1415907},
  number    = {525},
  pages     = {445-452},
  url       = {https://doi.org/10.1080/01621459.2017.1415907},
  volume    = {114},
  journal   = {J. Amer. Statist. Assoc.},
  publisher = {Taylor \& Francis},
  year      = {2019},
}

@Article{gneiting07_stric_proper_scorin_rules_predic_estim,
  author    = {Tilmann Gneiting and Adrian E Raftery},
  title     = {Strictly Proper Scoring Rules, Prediction, and Estimation},
  doi       = {10.1198/016214506000001437},
  number    = {477},
  pages     = {359-378},
  url       = {https://doi.org/10.1198/016214506000001437},
  volume    = {102},
  journal   = {J. Amer. Statist. Assoc.},
  publisher = {Taylor \& Francis},
  year      = {2007},
}

@article{friederichs12_forec_verif_extrem_value_distr,
  author =	 {Friederichs, Petra and Thorarinsdottir, Thordis L.},
  title =	 {Forecast Verification for Extreme Value Distributions
                  With an Application To Probabilistic Peak Wind
                  Prediction},
  journal =	 {Environmetrics},
  volume =	 23,
  number =	 7,
  pages =	 {579-594},
  year =	 2012,
  doi =		 {10.1002/env.2176},
  url =		 {https://doi.org/10.1002/env.2176},
}

@article{geirsson15_comput_effic_spatial_model_annual,
  author =	 {Geirsson, {\'O}li P. and Hrafnkelsson, Birgir and Simpson, Daniel},
  title =	 {Computationally Efficient Spatial Modeling of Annual Maximum 24-h Precipitation on
                  a Fine Grid},
  journal =	 {Environmetrics},
  volume =	 26,
  number =	 5,
  pages =	 {339-353},
  year =	 2015,
  doi =		 {10.1002/env.2343},
  url =		 {https://doi.org/10.1002/env.2343},
}

@Article{cooley07_bayes_spatial_model_extrem_precip_retur_level,
  author    = {Daniel Cooley and Douglas Nychka and Philippe Naveau},
  title     = {{B}ayesian Spatial Modeling of Extreme Precipitation Return Levels},
  doi       = {10.1198/016214506000000780},
  number    = {479},
  pages     = {824-840},
  url       = {https://doi.org/10.1198/016214506000000780},
  volume    = {102},
  journal   = {J. Amer. Statist. Assoc.},
  publisher = {Taylor \& Francis},
  year      = {2007},
}

@article{lehmann16_spatial_model_framew_charac_rainf,
  author =	 {Lehmann, Eric A. and Phatak, Aloke and Stephenson, Alec and Lau, Rex},
  title =	 {Spatial Modelling Framework for the Characterisation of Rainfall Extremes At
                  Different Durations and Under Climate Change},
  journal =	 {Environmetrics},
  volume =	 27,
  number =	 4,
  pages =	 {239-251},
  year =	 2016,
  doi =		 {10.1002/env.2389},
  url =		 {https://doi.org/10.1002/env.2389},
}

@Article{dyrrdal16_estim_extrem_areal_precip_norway,
  author    = {Anita Verpe Dyrrdal and Thomas Skaugen and Frode Stordal and Eirik J. F{\o}rland},
  title     = {Estimating Extreme Areal Precipitation in {N}orway From a Gridded Dataset},
  doi       = {10.1080/02626667.2014.947289},
  number    = {3},
  pages     = {483-494},
  url       = {https://doi.org/10.1080/02626667.2014.947289},
  volume    = {61},
  journal   = {Hydrol. Sci. J.},
  publisher = {Taylor \& Francis},
  year      = {2016},
}

@Article{barbero19_synth_hourl_daily_precip_extrem,
  author  = {Renaud Barbero and Hayley J. Fowler and Stephen Blenkinsop and Seth Westra and Vincent Moron and Elizabeth Lewis and Steven Chan and Geert Lenderink and Elizabeth Kendon and Selma Guerreiro and Xiao-Feng Li and Roberto Villalobos and Haider Ali and Vimal Mishra},
  title   = {A Synthesis of Hourly and Daily Precipitation Extremes in Different Climatic Regions},
  doi     = {10.1016/j.wace.2019.100219},
  pages   = {100219},
  url     = {https://doi.org/10.1016/j.wace.2019.100219},
  volume  = {26},
  journal = {Weather Clim. Extremes},
  year    = {2019},
}

@Article{sang10_contin_spatial_proces_model_spatial_extrem_values,
  author  = {Sang, Huiyan and Gelfand, Alan},
  title   = {Continuous Spatial Process Models for Spatial Extreme Values},
  doi     = {10.1007/s13253-009-0010-1},
  pages   = {49-65},
  url     = {https://doi.org/10.1007/s13253-009-0010-1},
  volume  = {15},
  journal = {J. Agric. Biol. Environ. Stat.},
  month   = {03},
  year    = {2010},
}

@article{lerch17_forec_dilem,
  author =	 "Lerch, Sebastian and Thorarinsdottir, Thordis L. and Ravazzolo, Francesco and
                  Gneiting, Tilmann",
  title =	 {Forecaster's Dilemma: Extreme Events and Forecast Evaluation},
  journal =	 "Statist. Sci.",
  volume =	 32,
  number =	 1,
  pages =	 "106--127",
  year =	 2017,
  doi =		 "10.1214/16-STS588",
  url =		 {https://doi.org/10.1214/16-STS588},
  fjournal =	 "Statistical Science",
  month =	 02,
  publisher =	 "The Institute of Mathematical Statistics",
}

@Article{gneiting11_compar_densit_forec_using_thres,
  author    = {Tilmann Gneiting and Roopesh Ranjan},
  title     = {Comparing Density Forecasts Using Threshold- and Quantile-Weighted Scoring Rules},
  doi       = {10.1198/jbes.2010.08110},
  number    = {3},
  pages     = {411-422},
  url       = {https://doi.org/10.1198/jbes.2010.08110},
  volume    = {29},
  journal   = {J. Bus. Econ. Stat.},
  publisher = {Taylor \& Francis},
  year      = {2011},
}

@article{galassi07_gnu_scien_librar_refer_manual,
  author =	 {Galassi, Mark and Davies, Jim and Theiler, James and Gough, Brian and Jungman,
                  Gerard and Alken, Patrick and Booth, Michael and Rossi, Fabrice and Ulerich, R},
  title =	 {The {GNU} Scientific Library Reference Manual, 2007},
  year =	 2007,
  URL =		 {http://www.gnu.org/software/gsl},
}

@Article{castro-camilo19_splic_gamma_gener_paret_model,
  author        = {Castro-Camilo, Daniela and Huser, Rapha{\"e}l and Rue, H{\aa}vard},
  title         = {A Spliced Gamma-Generalized {P}areto Model for Short-Term Extreme Wind Speed Probabilistic Forecasting},
  doi           = {10.1007/s13253-019-00369-z},
  number        = {3},
  pages         = {517--534},
  url           = {https://doi.org/10.1007/s13253-019-00369-z},
  volume        = {24},
  da            = {2019/09/01},
  id            = {Castro-Camilo2019},
  journal       = {JABES},
  ty            = {JOUR},
  year          = {2019},
}

@Manual{quantreg,
  author =	 {Roger Koenker},
  note =	 {R package version 5.75},
  title =	 {quantreg: Quantile Regression},
  url =		 {https://CRAN.R-project.org/package=quantreg},
  year =	 2020,
}

@Article{hankin06_special_funct_r,
  author =	 {Robin K. S. Hankin},
  title =	 {Special Functions in {R}: Introducing the {G}sl Package},
  journal =	 {R News},
  volume =	 6,
  year =	 2006,
  issue =	 4,
}

@article{mohr09_compar_version,
  author =	 {Mohr, Matthias},
  title =	 {Comparison of Versions 1.1 and 1.0 of Gridded Temperature and Precipitation Data
                  for {N}orway},
  journal =	 {met.no note},
  volume =	 19,
  year =	 2009,
}

@Article{tveito05_gis_based_agro_ecolog_decis,
  author  = {Tveito, Ole Einar and Bj{\o}rdal, Inge and Skjelv{\aa}g, Arne Oddvar and Aune, Bj{\o}rn},
  title   = {A {GIS}-Based Agro-Ecological Decision System Based on Gridded Climatology},
  doi     = {10.1017/S1350482705001490},
  eprint  = {https://rmets.onlinelibrary.wiley.com/doi/pdf/10.1017/S1350482705001490},
  number  = {1},
  pages   = {57-68},
  url     = {https://doi.org/10.1017/S1350482705001490},
  volume  = {12},
  journal = {Meteorol Appl},
  year    = {2005},
}

@Article{van12_spatial_regres_model_extrem_precip_belgium,
  author  = {Van de Vyver, H.},
  title   = {Spatial Regression Models for Extreme Precipitation in {B}elgium},
  doi     = {10.1029/2011WR011707},
  eprint  = {https://agupubs.onlinelibrary.wiley.com/doi/pdf/10.1029/2011WR011707},
  number  = {9},
  url     = {https://doi.org/10.1029/2011WR011707},
  volume  = {48},
  journal = {Water Resour. Res.},
  year    = {2012},
}

@Article{papalexiou13_battl_extrem_value_distr,
  author  = {Papalexiou, Simon Michael and Koutsoyiannis, Demetris},
  title   = {Battle of Extreme Value Distributions: a Global Survey on Extreme Daily Rainfall},
  doi     = {10.1029/2012WR012557},
  eprint  = {https://agupubs.onlinelibrary.wiley.com/doi/pdf/10.1029/2012WR012557},
  number  = {1},
  pages   = {187-201},
  url     = {https://doi.org/10.1029/2012WR012557},
  volume  = {49},
  journal = {Water Resour. Res.},
  year    = {2013},
}

@Article{castro-camilo20_local_likel_estim_compl_tail,
  author    = {Daniela Castro-Camilo and Rapha{\"e}l Huser},
  title     = {Local Likelihood Estimation of Complex Tail Dependence Structures, Applied To {U.S.} Precipitation Extremes},
  doi       = {10.1080/01621459.2019.1647842},
  eprint    = {https://doi.org/10.1080/01621459.2019.1647842},
  number    = {531},
  pages     = {1037-1054},
  url       = {https://doi.org/10.1080/01621459.2019.1647842},
  volume    = {115},
  journal   = {J. Amer. Statist. Assoc.},
  publisher = {Taylor \& Francis},
  year      = {2020},
}

@Article{matheson76_scorin_rules_contin_probab_distr,
  author  = {Matheson, James E. and Winkler, Robert L.},
  title   = {Scoring Rules for Continuous Probability Distributions},
  doi     = {10.1287/mnsc.22.10.1087},
  number  = {10},
  pages   = {1087-1096},
  url     = {https://doi.org/10.1287/mnsc.22.10.1087},
  volume  = {22},
  journal = {Manage Sci},
  year    = {1976},
}

@Article{sang09_hierar_model_extrem_values_obser,
  author  = {Sang, Huiyan and Gelfand, Alan E.},
  title   = {Hierarchical Modeling for Extreme Values Observed Over Space and Time},
  doi     = {10.1007/s10651-007-0078-0},
  number  = {3},
  pages   = {407--426},
  url     = {https://doi.org/10.1007/s10651-007-0078-0},
  volume  = {16},
  journal = {Environ. Ecol. Stat.},
  year    = {2009},
}

@Article{katz02_statis_extrem_hydrol,
  author  = {Richard W Katz and Marc B Parlange and Philippe Naveau},
  title   = {Statistics of Extremes in Hydrology},
  doi     = {10.1016/S0309-1708(02)00056-8},
  number  = {8},
  pages   = {1287-1304},
  url     = {https://doi.org/10.1016/S0309-1708(02)00056-8},
  volume  = {25},
  journal = {Adv. Water Res.},
  year    = {2002},
}

@Article{wilson05_fundam_probab_distr_heavy_rainf,
  author  = {Wilson, P. S. and Toumi, R.},
  title   = {A Fundamental Probability Distribution for Heavy Rainfall},
  doi     = {10.1029/2005GL022465},
  number  = {14},
  url     = {https://doi.org/10.1029/2005GL022465},
  volume  = {32},
  journal = {Geophys. Res. Lett.},
  year    = {2005},
}

@book{stein99_inter,
  author =	 {Stein, Michael L},
  title =	 {Interpolation of spatial data : some theory for {K}riging},
  year =	 1999,
  publisher =	 {Springer},
  address =	 {New York},
  isbn =	 0387986294,
  series =	 {Springer series in statistics},
}

@book{matern86_spatial_variat,
  author =	 {Matern, B},
  title =	 {Spatial Variation},
  year =	 1986,
  publisher =	 {Springer New York : Imprint: Springer},
  url =		 {https://doi.org/10.1007/978-1-4615-7892-5},
  address =	 {New York, NY},
  doi =		 {10.1007/978-1-4615-7892-5},
  edition =	 {2nd ed. 1986.},
  isbn =	 {1-4615-7892-2},
  series =	 {Lecture Notes in Statistics},
  volume =	 36,
}

@article{guttorp06_studies_histor_probab_statis_xlix,
  author =	 {Guttorp, Peter and Gneiting, Tilmann},
  title =	 {{Studies in the History of Probability and Statistics XLIX on the Mat{\'e}rn
                  Correlation family}},
  journal =	 {Biometrika},
  volume =	 93,
  number =	 4,
  pages =	 {989-995},
  year =	 2006,
  doi =		 {10.1093/biomet/93.4.989},
  url =		 {https://doi.org/10.1093/biomet/93.4.989},
  eprint =	 {https://academic.oup.com/biomet/article-pdf/93/4/989/593552/934989.pdf},
  issn =	 {0006-3444},
  month =	 12,
}

@article{bolin19_scale_depen,
  author =	 {Bolin, David and Wallin, Jonas},
  title =	 {Scale Dependence: Why the Average {CRPS} Often Is Inappropriate for Ranking
                  Probabilistic Forecasts},
  journal =	 {arXiv:1912.05642},
  year =	 2019,
}

@report{wef_2021,
  author = {{World Economic Forum}},
  title = {The Global Risks Report 2021},
  year = 2021,
  url = {http://www3.weforum.org/docs/WEF_The_Global_Risks_Report_2021.pdf},
}

@Article{sun15_mater_model_spatial_covar_struc,
  author  = {Sun, Ying and Bowman, Kenneth P. and Genton, Marc G. and Tokay, Ali},
  title   = {A {M}at{\'e}rn Model of the Spatial Covariance Structure of Point Rain Rates},
  doi     = {10.1007/s00477-014-0923-2},
  number  = {2},
  pages   = {411--416},
  url     = {https://doi.org/10.1007/s00477-014-0923-2},
  volume  = {29},
  journal = {Stoch Env Res Risk A},
  year    = {2015},
}

@article{hanssen-bauer98_annual_seasonal_prec_var,
  author = {Hanssen-Bauer, I. and F{\o}rland, E. J.},
  title = {Annual and seasonal precipitation variations in {N}orway 1896-1997},
  year = 1998,
  journal = {{DNMI KLIMA} Report 27/98},
}

@misc{castro-camilo21_pract_strat_gev_regres_model_extrem,
  author =	 {Castro-Camilo, Daniela and Huser, Rapha{\"e}l and Rue, H{\aa}vard},
  title =	 {Practical Strategies for {GEV}-based Regression Models for Extremes},
  year =	 2021,
  doi =		 {10.48550/ARXIV.2106.13110},
  url =		 {https://arxiv.org/abs/2106.13110},
  publisher =	 {arXiv},
  copyright =	 {Creative Commons Attribution 4.0 International}
}

@article{robinson00_extrem_analy_proces_sampl_at_differ_frequen,
  author =	 {Robinson, M. E. and Tawn, J. A.},
  title =	 {Extremal Analysis of Processes Sampled At Different Frequencies},
  journal =	 {J. R. Stat. Soc. B},
  volume =	 62,
  number =	 1,
  pages =	 {117-135},
  year =	 2000,
  doi =		 {10.1111/1467-9868.00223},
  url =		 {https://doi.org/10.1111/1467-9868.00223},
}

@article{buecher17_maxim_likel_estim_gener_extrem_value_distr,
  author =	 {B{\"u}cher, Axel and Segers, Johan},
  title =	 {On the Maximum Likelihood Estimator for the Generalized Extreme-Value
                  Distribution},
  journal =	 {Extremes},
  volume =	 20,
  number =	 4,
  pages =	 {839--872},
  year =	 2017,
  doi =		 {10.1007/s10687-017-0292-6},
  url =		 {https://doi.org/10.1007/s10687-017-0292-6},
}

@article{zou19_multip_block_sizes_overl_block,
  author =	 {Nan Zou and Stanislav Volgushev and Axel B{\"u}cher},
  title =	 {Multiple Block Sizes and Overlapping Blocks for Multivariate Time Series Extremes},
  journal =	 {arXiv:1907.09477},
  year =	 2019,
  archivePrefix ={arXiv},
  eprint =	 {1907.09477},
  primaryClass = {math.ST},
}

@article{buecher18_infer_heavy_tailed_station_time,
  author =	 {Axel B{\"u}cher and Johan Segers},
  title =	 {Inference for Heavy Tailed Stationary Time Series Based on Sliding Blocks},
  journal =	 {Electronic Journal of Statistics},
  volume =	 12,
  number =	 1,
  pages =	 {1098 -- 1125},
  year =	 2018,
  doi =		 {10.1214/18-EJS1415},
  url =		 {https://doi.org/10.1214/18-EJS1415},
  publisher =	 {Institute of Mathematical Statistics and Bernoulli Society},
}

@inproceedings{vandeskog21_model_block_maxim,
  author =	 {Silius M. Vandeskog and Sara Martino and Daniela Castro-Camilo},
  title =	 {Modelling Block Maxima with the blended generalised extreme value distribution},
  booktitle =	 {22nd {E}uropean {Y}oung {S}tatisticians {M}eeting - {P}roceedings},
  year =	 2021,
  isbn =	 {978-960-7943-23-1},
}

@article{ulrich20_estim_idf_curves_consis_over,
  author =	 {Ulrich, Jana and Jurado, Oscar E. and Peter, Madlen and Scheibel, Marc and Rust,
                  Henning W.},
  title =	 {Estimating {IDF} Curves Consistently Over Durations With Spatial Covariates},
  journal =	 {Water},
  volume =	 12,
  number =	 11,
  year =	 2020,
  doi =		 {10.3390/w12113119},
  url =		 {https://doi.org/10.3390/w12113119},
  issn =	 {2073-4441},
}

@book{dalrymple60_flood,
  author =	 {Dalrymple, Tate},
  title =	 {Flood-frequency analyses},
  year =	 1960,
  publisher =	 {U.S. Government Printing Office},
  number =	 1543,
}

@article{wang16_bayes_hierar_model_spatial_extrem,
  author =	 {Yixin Wang and Mike K.P. So},
  title =	 {A {B}ayesian Hierarchical Model for Spatial Extremes With Multiple Durations},
  journal =	 {Computational Statistics \& Data Analysis},
  volume =	 95,
  pages =	 {39-56},
  year =	 2016,
  doi =		 {10.1016/j.csda.2015.09.001},
  url =		 {https://doi.org/10.1016/j.csda.2015.09.001},
  issn =	 {0167-9473},
}

@article{koutsoyiannis98_mathem_framew_study_rainf_inten,
  author =	 {Demetris Koutsoyiannis and Demosthenes Kozonis and Alexandros Manetas},
  title =	 {A Mathematical Framework for Studying Rainfall Intensity-Duration-Frequency
                  Relationships},
  journal =	 {Journal of Hydrology},
  volume =	 206,
  number =	 1,
  pages =	 {118-135},
  year =	 1998,
  doi =		 {10.1016/S0022-1694(98)00097-3},
  url =		 {https://doi.org/10.1016/S0022-1694(98)00097-3},
  issn =	 {0022-1694},
}

@book{hosking97_region,
  author =	 {Hosking, Jonathan Richard Morley and Wallis, James R},
  title =	 {Regional frequency analysis},
  year =	 {1997},
}

@article{carreau16_charac,
  author =	 {Carreau, Julie and Naveau, Philippe and Neppel, Luc},
  note =	 {working paper or preprint},
  title =	 {{Characterization of homogeneous regions for regional peaks-over-threshold
                  modeling of heavy precipitation}},
  url =		 {https://hal.ird.fr/ird-01331374},
  year =	 2016,
}

@article{naveau14_fast_nonpar_spatio_tempor_regres,
  author =	 {Naveau, P. and Toreti, A. and Smith, I. and Xoplaki, E.},
  title =	 {A Fast Nonparametric Spatio-Temporal Regression Scheme for Generalized {P}areto
                  Distributed Heavy Precipitation},
  journal =	 {Water Resources Research},
  volume =	 50,
  number =	 5,
  pages =	 {4011-4017},
  year =	 2014,
  doi =		 {10.1002/2014WR015431},
  url =		 {https://doi.org/10.1002/2014WR015431},
}

@article{whittle54_station_proces_plane,
  author =	 {P. Whittle},
  title =	 {On Stationary Processes in the Plane},
  journal =	 {Biometrika},
  volume =	 41,
  number =	 {3/4},
  pages =	 {434--449},
  year =	 1954,
  ISSN =	 00063444,
  URL =		 {http://www.jstor.org/stable/2332724},
  publisher =	 {[Oxford University Press, Biometrika Trust]},
}

\appendix

\renewcommand{\thesection}{S\arabic{section}}
\numberwithin{equation}{section} 
\numberwithin{figure}{section} 

\section*{\huge Supplementary material}

\section{Simulation study}

A simulation study is conducted for comparing the performance of the generalised extreme value (GEV) distribution
and blended generalised extreme value (bGEV) distribution in a
univariate setting. Our penultimate goal when modelling block maxima is to estimate return
levels. Thus, the two distributions are compared by evaluating the performance of their return
level estimators.

We sample \(n \in \{25, 50, 100, 500, 1000\}\) i.i.d.\ ``block maxima'' from a GEV distribution resembling
the estimated distribution of the yearly maxima of hourly precipitation. From Table~3 and Table~4 in
``Modelling sub-daily precipitation extremes with the blended generalised extreme value
distribution'' we find that the posterior mean of the intercept in \(\bm \beta_\mu\) is
\(E[\beta_{\mu, 0}] = 0.661\), and the posterior mean of the intercept in \(\bm \beta_\sigma\) is
\(E[\beta_{\sigma, 0}] = 2.835\). Consequently, we set \(\alpha = 0.5\), \(\beta = 0.8\),
\(\mu_\alpha = E[\beta_{\mu, 0}] \cdot \exp(E[\beta_{\sigma, 0}]) \approx 11.26\), and
\(\sigma_\beta = E[\sigma_\beta^*] \cdot \exp(E[\beta_{\sigma, 0}]) \approx 2.01\). The tail
parameter is set equal to the posterior mean from Table~4: \(\xi = 0.178\). Using the \(n\) block
maxima, we compute maximum likelihood estimators for the parameters of the GEV and bGEV
distributions, which are further used for computing return level estimators for periods of length
25, 50, 100, 250 and 500. The actual maximum likelihood estimation is performed on the \((\mu,
\sigma, \xi)\) parametrisation, in which \(\mu \approx 10.05\) and \(\sigma \approx
3.21\). Optimisation is performed using the \texttt{optim} function in \texttt{R}, where the true
values of \((\mu, \sigma, \xi)\) are used as initial values. The entire procedure is repeated 500
times for each value of \(n\). This allows us to estimate the distribution of the maximum likelihood
estimators for all return levels in question.

\begin{figure}
  \centering
  \includegraphics[width=.98\linewidth,page=1]{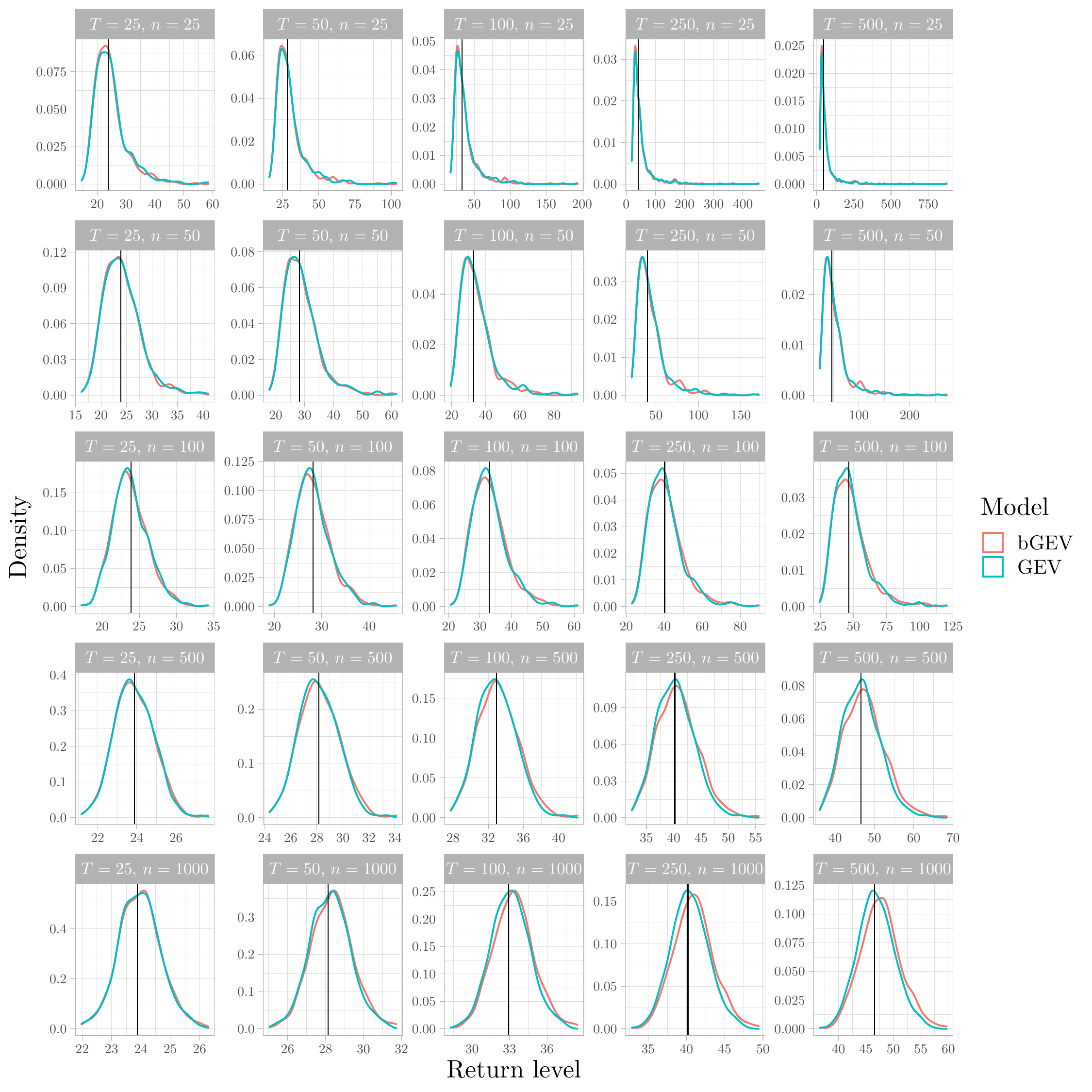}
  \caption{Distributions of return level estimators computed from model fits with the GEV and bGEV
    distributions to simulated data from a GEV distribution. The dark vertical lines display the true return
    levels. Return periods are denoted as \(T\), with unit ``block sizes'', and the number of
    block maxima used for fitting the GEV and bGEV distributions is denoted as \(n\).}
  \label{fig:simulation1}
\end{figure}

Figure~\ref{fig:simulation1} displays the distribution of all return level estimators using both the
GEV and the bGEV distributions. The difference between the distributions of the GEV estimators and
the bGEV estimators are negligible for small values of \(n\). For large return periods and values of
\(n\) some slight differences are appearing. However, in practical situations one rarely encounters
as much as 1000 block maxima, and the blocks are almost never large enough for the limit
distribution to hold exactly. Thus, the differences in distribution can be considered negligible for large
values of \(n\) for almost all real-life settings.
Larger differences in model fit could theoretically occur for small blocks where the block maxima
deviate further from the GEV distribution. However, in such settings, both the GEV and the bGEV
distribution would be misspecified, and we have no reason to believe that the misspecification in
the GEV distribution would be strictly better or worse than the misspecification in the bGEV distribution.
Therefore, our results imply that we do not lose anything in performance by modelling
GEV data with the bGEV distribution instead of the GEV distribution.

\begin{figure}
  \centering
  \includegraphics[width=.98\linewidth,page=2]{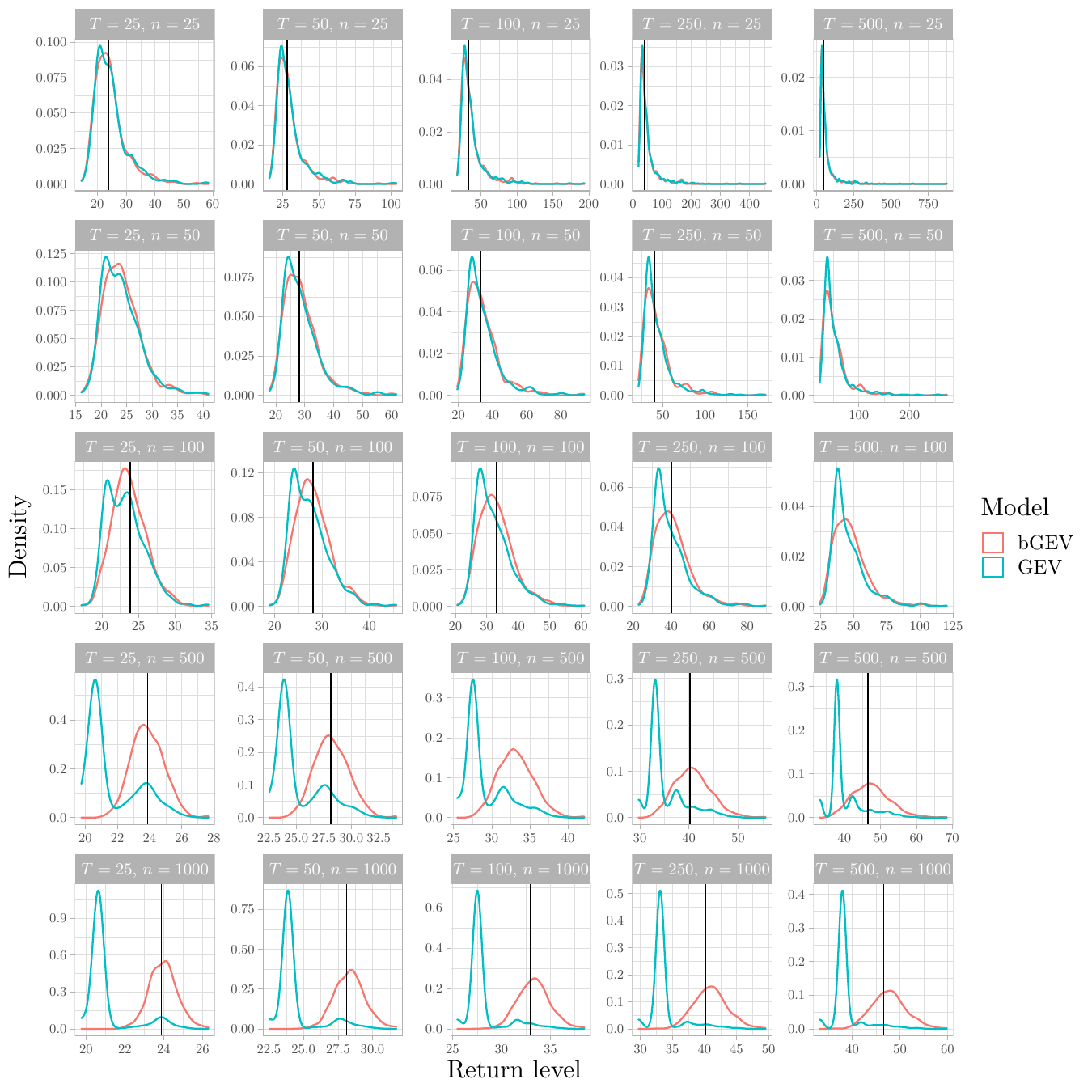}
  \caption{
    Distributions of return level estimators computed from model fits with the GEV and bGEV
    distributions to simulated data from a GEV distribution, when using bad initial values. The dark vertical lines display the true return
    levels. Return periods are denoted as \(T\), with unit ``block sizes'', and the number of
    block maxima used for fitting the GEV and bGEV distributions is denoted as \(n\).}
  \label{fig:simulation2}
\end{figure}

In most real-life settings we do not have as much prior knowledge about the true underlying
distribution, and we will likely choose less optimal initial values and/or prior distributions.
To examine the effects of choosing less optimal initial values, we repeat the maximum likelihood estimation on the
exact same data used for Figure~\ref{fig:simulation1}, using different initial values. Figure~\ref{fig:simulation2}
displays the distributions of the return level estimators when we use an initial value of \(0.9\)
instead of \(3.2\) for \(\sigma\), without changing the initial values for \(\mu\) or \(\xi\). For
small values of \(n\)
the two distributions yield comparable results. However, for large values of \(n\), the performance
of the GEV distribution quickly deteriorates, and the return level estimators are shifted to the
left. The bGEV estimators, however, seem to have identical distributions in
Figures~\ref{fig:simulation1} and~\ref{fig:simulation2}, and are not affected by the slight change
of initial values. This demonstrates how inference with the
bGEV distribution is more robust than inference with 
the GEV distribution. The parameter-dependent support of the GEV complicates likelihood-based
inference and makes it more difficult to locate the global maximum of the likelihood, even in simple
and univariate settings with unrealistically large amounts of data that are perfectly
GEV-distributed.

In such a simple setting, it is relatively easy to see that we chose bad initial
values for \((\mu, \sigma, \xi)\). However, when modelling block maxima with a large number of
covariates and in high-dimensional settings, it becomes considerably more difficult to choose good
enough initial values and/or prior distributions. Additionally, it is not trivial to examine a
maximum likelihood estimator or 
posterior distribution and find out if the inference procedure was successful or not. The GEV return
level estimators in Figure~\ref{fig:simulation2} take on reasonable values and are large enough to
seem realistic. Without knowing the truth, it would not be trivial for a practitioner to detect that
they are faulty, and caused by numerical
problems. Thus, if one performs likelihood-based inference using the GEV distribution,
one should always perform extensive checks to evaluate whether the fit is reasonable, or if it
is caused by numerical problems. The increased robustness of the bGEV distribution provides
more safety, and most practitioners can be more confident in their model fits after fitting the
bGEV distribution to block maxima data instead of the GEV distribution.

Similarly, while attempting to model return levels for precipitation in Norway using the \texttt{gev} family in
\texttt{R-INLA}, we experienced that e.g.\ adding or removing one covariate from the model could
completely alter the model fits if our priors were too wide or the data were not correctly
standardised. Examining which of the fits were the best required comprehensive and time-demanding
cross-validation studies, and it was difficult to find out if the best model fit was ``good
enough'', or if a slight change in priors or covariates would yield considerably better model
fits. These problems were considerably mitigated by switching to the \texttt{bgev} family, which
resulted in more stable inference using less informative priors, and less optimal initial values.

\section{PC prior for the tail parameter}

In order to compute the PC prior for the GEV and bGEV distribution with \(\xi \geq 0\) and base model \(\xi = 0\), one must first
compute KLD\((\pi_\xi, \pi_0)\) for the GEV distribution and the bGEV distribution. Notice that
the base model is identical for both distributions.
Writing the GEV distribution function as \(F(x) = \exp(-h(x))\) 
gives the probability density function 
\[
\pi_\xi(x) = - \exp(-h(x)) h'(x), \quad
h(x) = (1 + \xi (x - \mu) / \sigma)^{-1 / \xi}_+.
\]
The KLD for the GEV distribution is equal to
\begin{linenomath*}
\begin{equation*}
  \label{eq:kld-gev-1}
  \begin{aligned}
  \text{KLD}(\pi_\xi, \pi_0) =& \int \pi_\xi(x) \log\left(\frac{\pi_\xi(x)}{\pi_0(x)}\right)
  \text{d}x \\
  =& \int - e^{-h(x)} h'(x) \cdot \left( -h(x) + \log(- \sigma h'(x))
    + \exp\left(-\frac{x -
        \mu}{\sigma}\right) + \frac{x - \mu}{\sigma} \right) \text{d}x.
  \end{aligned}
\end{equation*}
\end{linenomath*}
The fourth term of the KLD is simply equal to the expectation of \((x - \mu) / \sigma\), which is known. 
The first term is easily solvable with the substitution \(u = h(x)\). Using
the same substitution for the second term with the knowledge that \(\log(- \sigma h'(x)) = (1 + \xi)
\log(h(x))\) gives the integral
\begin{linenomath*}
\begin{equation*}
  \int - e^{-h(x)} h'(x) \log(- \sigma h'(x)) = \int_0^\infty e^{-u} (1 + \xi) \log (u)\  \text{d}u
  = - (1 + \xi) \gamma,
\end{equation*}
\end{linenomath*}
where \(\gamma\) is the Euler-Mascheroni constant. We are unable to find a closed-form expression for
the third term. However, using substitution with \(u = h(x)\) gives
\begin{linenomath*}
\begin{equation*}
    \int - e^{-h(x)} h'(x) \exp\left(-\frac{x - \mu}{\sigma}\right) \text{d}x
    =\int_0^\infty
    \exp\left(\frac{1}{\xi} - u - \frac{u^{-\xi}}{\xi}\right) \text{d}u,
\end{equation*}
\end{linenomath*}
which is finite for \(\xi > 0\). With a change of variables, this integral is transformed to have finite limits,
\begin{linenomath*}
\begin{equation*}
  \label{eq:integral-transform}
  \int_0^\infty \exp\left(\frac{1}{\xi} - \left(u + u^{-\xi}/\xi\right)\right) \text{d}u
  =\int_0^1 \exp\left(\frac{1}{\xi} \left(1 - (-\log v)^{-\xi}\right)\right) \text{d}v.
\end{equation*}
\end{linenomath*}
This can easily be numerically approximated.
Summarising, the KLD for the GEV is finite for \(0 \leq \xi < 1\) and equal to
\begin{equation}
  \label{eq:gev-kld}
  \text{KLD}(\pi_\xi, \pi_0) = -1 - (1 + \xi) \gamma + \xi^{-1}(\Gamma(1 - \xi) - 1) +
  \int_0^1 \exp\left(\frac{1}{\xi} \left(1 - (-\log v)^{-\xi}\right)\right) \text{d}v.
\end{equation}

The PC prior has probability density function \(\pi(d) = \lambda \exp(-\lambda d)\)
with \(d = \sqrt{2 \text{KLD}(\pi_\xi, \pi_0)}\).
Transforming the PC prior from a distribution on \(d\) to a distribution on \(\xi\) gives
\begin{equation}
  \label{eq:pc-prior-xi}
  \pi(\xi) = \lambda \exp\left(-\lambda d(\xi)\right) \left| \frac{\partial
      d(\xi)}{\partial \xi} \right|
  = \frac{\lambda}{d(\xi)} \exp\left(-\lambda d(\xi)\right) \left|
    \frac{\partial}{\partial \xi} \text{KLD}(\pi_\xi, \pi_0)\right|.
\end{equation}
Consequently, the derivative of the KLD must also be computed.
Using derivation under the integral sign gives
\begin{linenomath*}
\begin{equation*}
  \label{eq:gev-kld-derivative}
  \frac{\partial}{\partial \xi} \text{KLD}(\pi_\xi, \pi_0) = -\gamma - \frac{\Gamma(1 - \xi)
    \Psi(1 - \xi)}{\xi} - \frac{\Gamma(1 - \xi) - 1}{\xi^2} + \int_0^1 g(v; \xi) \text{d}v,
\end{equation*}
\end{linenomath*}
where \(\Psi(\cdot)\) is the digamma function, and
\begin{linenomath*}
\begin{equation*}
  g(v; \xi) = \exp\left(\frac{1}{\xi} \left(1 - (-\log v)^{-\xi}\right)\right)
  \frac{1}{\xi^2} \left(-1 + (-\log v)^{-\xi} \left(1 + \xi \log(-\log v)\right)\right).
\end{equation*}
\end{linenomath*}
This integral must also be numerically approximated.

When computing the KLD for the bGEV distribution, the integral for the KLD is divided into
three parts:
\begin{linenomath*}
\begin{equation*}
  \label{eq:bgev-kld}
  \begin{aligned}
  &\text{KLD}(\pi_\xi, \pi_0) = \\
  &\qquad \int_{-\infty}^{p_a} \pi_\xi(x) \log \left(\frac{\pi_\xi(x)}{\pi_0(x)}\right) \text{d}x
  +
  \int_{p_a}^{p_b} \pi_\xi(x) \log \left(\frac{\pi_\xi(x)}{\pi_0(x)}\right) \text{d}x
  + \int_{p_b}^{\infty} \pi_\xi(x) \log \left(\frac{\pi_\xi(x)}{\pi_0(x)}\right) \text{d}x \\
  &\qquad= \text{KLD}_{\text{Gumbel}}(\pi_\xi, \pi_0)
  +
  \text{KLD}_{\text{Blending}}(\pi_\xi, \pi_0)
  + \text{KLD}_{\text{Fréchet}}(\pi_\xi, \pi_0).
  \end{aligned}
\end{equation*}
\end{linenomath*}
A closed-form expression can be found for the Gumbel part of the KLD, where we express the Gumbel
distribution function as \(G(x) = \exp(-h_2(x))\) and use the same substitution techniques as for the
KLD computations with the GEV distribution. We get
\begin{linenomath*}
\begin{equation*}
  \label{eq:bgev-kld-gumbel}
  \begin{aligned}
    &\text{KLD}_{\text{Gumbel}}(\pi_\xi, \pi_0) = \\
    &\qquad-\Gamma_u(-\log p_a; 2) + p_a \log(-\log p_a) - E_i(\log p_a)
    + \Gamma_u(-\log p_a; C_1 + 1) e^{-C_2} \\
    &\qquad- C_1 \left(p_a \log(-\log p_a) - E_i(\log p_a)\right)
    + p_a \left(C_2 + \log \xi +
    \log\left(\log\left(\frac{\log p_a}{\log p_b}\right)\right)\right) \\
  &\qquad- p_a\log\left((-\log p_b)^{-\xi} -
      (-\log p_a)^{-\xi}\right),
  \end{aligned}
\end{equation*}
\end{linenomath*}
where
\begin{linenomath*}
\begin{equation*}
  \label{eq:bgev-kld-gumbel-constants}
  C_1 = \frac{1}{\xi} \frac{(-\log p_b)^{-\xi} - (-\log p_a)^{-\xi}}{\log\left(\frac{\log p_a}{\log
        p_b}\right)} \quad \text{and} \quad
  C_2 = C_1 \log(-\log p_a) + \frac{1}{\xi} \left((-\log
    p_a)^{-\xi} - 1\right).
\end{equation*}
\end{linenomath*}
Here, \(\Gamma_u(x; \alpha) =
\int_x^\infty t^{\alpha - 1} e^{-t} \text{d}t\) is the upper incomplete gamma function, and \(E_i(x) = \int_{-\infty}^x (e^t/t)
\text{d}t\) is the exponential integral, which can be evaluated using the GNU
Scientific Library \citep{galassi07_gnu_scien_librar_refer_manual, hankin06_special_funct_r}.
The Fréchet part of the KLD is found in the same way as \eqref{eq:gev-kld}, only using different
integration limits. For \(0 \leq \xi < 1\) we get
\begin{linenomath*}
\begin{equation*}
  \label{eq:bgev-kld-frechet}
  \begin{aligned}
    \text{KLD}_{\text{Fréchet}}(\pi_\xi, \pi_0) =& -\Gamma_l(-\log p_b; 2) + (\xi + 1) \left(-p_b
      \log(-\log p_b) + E_i(\log p_b) - \gamma\right) \\
    &+ \frac{1}{\xi} \left(\Gamma_l(-\log p_b; 1 - \xi) - (1 - p_b)\right)
    + \int_{p_b}^1 \exp\left(\frac{1}{\xi} \left(1 - (-\log u)^{-\xi}\right)\right) \text{d}u,
  \end{aligned}
\end{equation*}
\end{linenomath*}
where \(\Gamma_l(x; \alpha) = \Gamma(\alpha) -
\Gamma_u(x; \alpha)\) is the lower incomplete gamma function. The integral in the expression above must once again be evaluated
numerically.
We are unable to find an expression for the KLD integral between the \(p_a\) quantile and the
\(p_b\) quantile. Thus, we use numerical integration with \(\mu = 0\) and \(\sigma = 1\) to compute
\(\text{KLD}_{\text{Blending}}(\pi_\xi, \pi_0)\). The value of the integral does not depend on the
values of \(\mu\) and \(\sigma\), but we are unable to compute it without choosing some value for
them. Being unable to find an expression for the KLD of the bGEV distribution, we must also use
numerical derivation to estimate the derivative of the KLD, needed in~\eqref{eq:pc-prior-xi}.

\begin{figure}
  \centering
  \includegraphics[width=.9\linewidth]{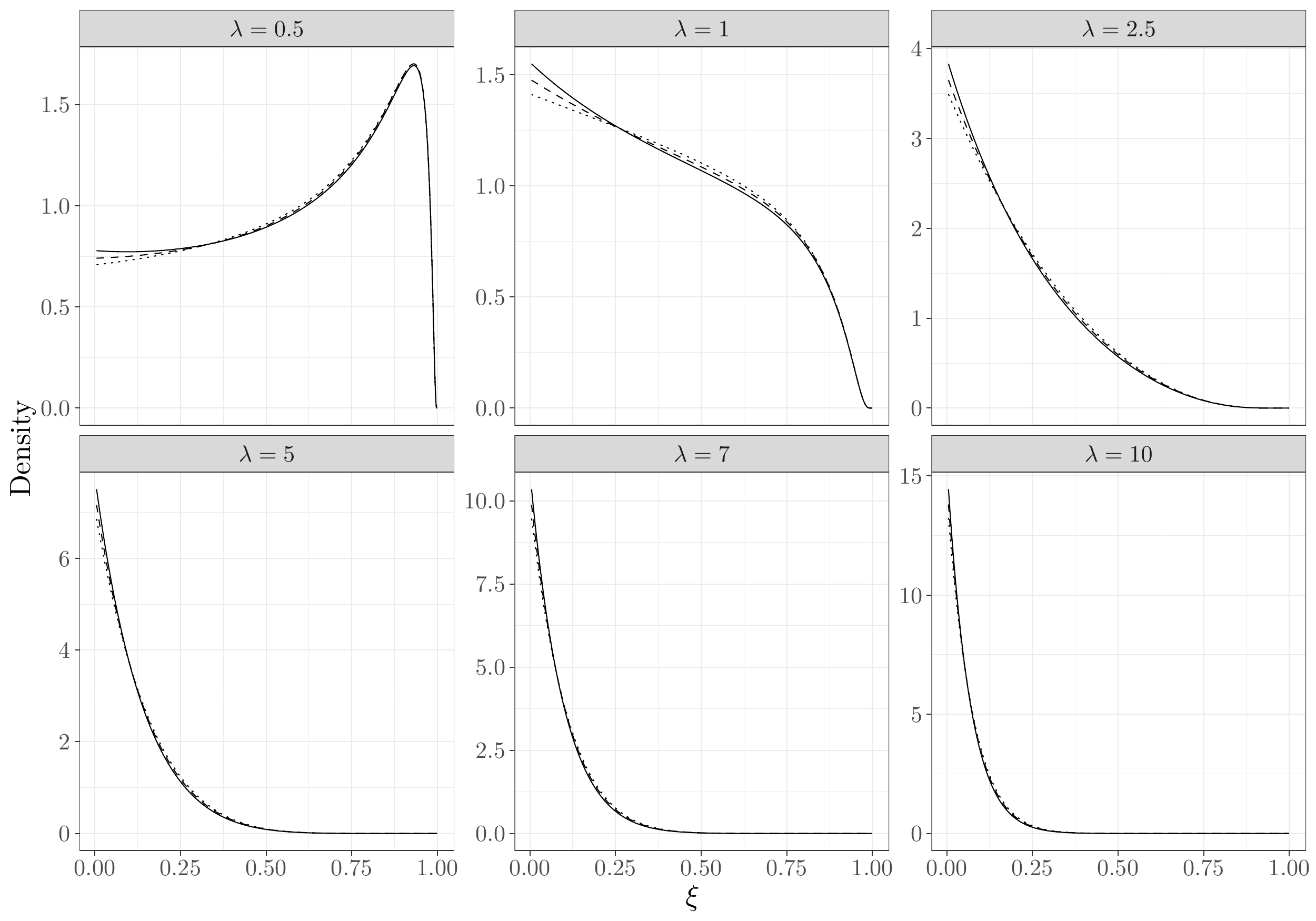}
  \caption{PC priors with base models \(\xi = 0\) and different penalty parameters \(\lambda\) for the GEV
    (solid line), bGEV (dashed line) and generalised Pareto distribution (dotted line).}
  \label{fig:pc-priors}
\end{figure}

The PC priors for the GEV, bGEV and generalised Pareto
distributions are displayed in Figure~\ref{fig:pc-priors} for different values of the penalty
parameter \(\lambda\). For all the chosen values of \(\lambda\) the three distributions are so
similar that their effect on a posterior distribution probably will be close to identical.
The PC prior for the generalised Pareto distribution exists
in closed-form and is already implemented in \texttt{R-INLA}, while the other PC priors must be computed
numerically and are not implemented in \texttt{R-INLA}. Consequently, we choose to use the PC prior
of the generalised Pareto
distribution for modelling the tail parameter of the bGEV distribution.

\end{document}